\documentclass[aps,prd,twocolumn,groupedaddress,amsmath,amssymb]{revtex4-1}
\usepackage{graphicx}  
\usepackage{dcolumn}   
\usepackage{bm}        
\usepackage{verbatim}   
\usepackage{subfigure}
\usepackage[utf8]{inputenc}
\usepackage{color}
\usepackage{braket}
\usepackage{slashed}
\usepackage{amssymb}

\begin{document}

\title{Production and polarization of direct $J/\psi$ to ${\mathcal O}(\alpha_s^3)$ in the improved color evaporation model in collinear factorization}
\author{Vincent Cheung}
\thanks{Present address: Nuclear and Chemical Sciences Division, Lawrence Livermore National Laboratory, Livermore, California, 94551, USA.}
\affiliation{
   Department of Physics and Astronomy,
   University of California, Davis,
   Davis, California 95616, USA
   }
\author{Ramona Vogt}
\affiliation{
   Nuclear and Chemical Sciences Division,
   Lawrence Livermore National Laboratory,
   Livermore, California 94551, USA
   }
\affiliation{
   Department of Physics and Astronomy,
   University of California, Davis,
   Davis, California 95616, USA
   }
\date{\today}
\begin{abstract}
We calculate the production and polarization of direct $J/\psi$ in the improved color evaporation model at $\mathcal{O}(\alpha_s^3)$ in the collinear factorization approach. We present the first calculation of polarization parameters $\lambda_\vartheta$, $\lambda_{\varphi}$, and $\lambda_{\vartheta \varphi}$ in the helicity and the Collins-Soper frames, as well as the frame-invariant polarization parameter $\tilde{\lambda}$ as a function of transverse momentum. We find agreement with both $J/\psi$ cross sections and the invariant polarization parameters at small and moderate $p_T$.
\end{abstract}


\maketitle


\section{Introduction}

Quarkonium production is important to understand both the long- and short-distance aspects of QCD. Both the perturbative and nonperturbative natures of QCD are needed to model the production of quarkonium from heavy quark production in hard processes to the hadronization of the final state. Nonrelativistic QCD (NRQCD) \cite{Caswell:1985ui}, the most commonly employed model of quarkonium production cannot describe the $J/\psi$ production and polarization while respecting the universality of the long-distance matrix elements for $p_T$ cuts less than twice the mass of the quarkonium state \cite{Bodwin:2014gia,Faccioli:2014cqa}. It also has difficulty describing the $J/\psi$ polarization and the LHCb $\eta_c$ production \cite{Han:2014jya,Zhang:2014ybe} while using heavy quark spin symmetry \cite{Neubert:1993mb,DeFazio:2000up,Casalbuoni:1996pg}. On the other hand, the color evaporation model (CEM) \cite{Barger:1979js,Barger:1980mg,Gavai:1994in} and the improved CEM (ICEM) \cite{Ma:2016exq} have only been employed extensively to hadroproduction on $S$-state quarkonia. 

Our previous charmonium and bottomonium polarization calculations performed in the $k_T$-factorization approach \cite{Cheung:2018tvq,Cheung:2018upe} describe both polarization and production at most $p_T$. However, the $p_T$ dependence of the production calculation has a strong dependence on the factorization scale chosen. Also, the discrepancies between the ICEM polarization calculation and the measured data are not visualized in a frame-independent way. We address these issues in this paper by performing a polarized production calculation in the collinear factorization approach and also computing a frame-invariant polarization parameter to compare with the data. In this calculation, only the production and polarization of direct $J/\psi$ is presented. We will address the effects of feed-down production on $J/\psi$ in a later publication.

In this paper, we present both the yield and the polarization parameters of direct $J/\psi$ production as a function of $p_T$ in the ICEM \cite{Ma:2016exq} using the collinear factorization approach. The polarized cross section calculation is presented in Sec.~\ref{section2}. The results, along with comparison of unpolarized $p_T$ distributions and polarization parameters to data, are shown in Sec.~\ref{section3}. Our conclusions are presented in Sec.~\ref{section4}.

\section{Polarized Cross Section}
\label{section2}
The ICEM assumes the $J/\psi$ production cross section takes a constant fraction of the open $c\bar{c}$ cross section with invariant mass above the mass of the $J/\psi$ but below the hadron threshold, the $D\overline{D}$ pair mass. A distinction is also made between the $c\bar{c}$ momentum and the $J/\psi$ momentum in the ICEM compared to the traditional CEM. The unpolarized direct $J/\psi$ production cross section in $p+p$ collision in the ICEM is given by
\begin{eqnarray}
\label{ch6-icem-cross-section}
\sigma &=& F_{J/\psi} \sum_{i,j}  \int^{2m_D}_{M_{J/\psi}}dM dx_i dx_j f_i(x_i,\mu_F)f_j(x_j,\mu_F) \nonumber \\
&\times& \hat{\sigma}_{ij\rightarrow c\bar{c}+k}(p_{c\bar{c}},\mu_R) |_{p_{c\bar{c}} = \frac{M}{M_{J/\psi}} p_{\psi}} \;,
\end{eqnarray}
where $i$ and $j$ are $q, \bar{q}$ and $g$ such that $ij = q\bar{q}$, $qg$, $\bar{q}g$ or $gg$, $F_{J/\psi}$ is a universal factor at fixed order for direct $J/\psi$ production in the ICEM independent of projectile and energy, $x$ is the momentum fraction of the parton, and $f(x,\mu_F)$ is the parton distribution function (PDF) for a parton in the proton as a function of $x$ and the factorization scale $\mu_F$. Finally, $\hat{\sigma}_{ij\rightarrow c\bar{c}+k}$ are the parton-level cross sections for initial states $ij$ to produce a $c\bar{c}$ pair with a light final-state parton $k$. The invariant mass of the $c\bar{c}$ pair, $M$, is integrated from the physical mass of $J/\psi$ ($M_{J/\psi} =3.10$~GeV) to two times the mass of the $D^0$ hadron ($2m_{D^0} = 3.72$~GeV). Because the $\mathcal{O}(\alpha_s^3)$ contribution diverges when the light parton is soft, in order to describe the $p_T$ distribution at low $J/\psi$ $p_T$, the initial-state partons are each given a small transverse momentum, $k_T$, kick of $\langle k_T^2\rangle = 1+ (1/12) \ln(\sqrt{s}/20 {\rm ~GeV})= 1.49{\rm ~GeV}^2$ for $\sqrt{s}=7$~TeV. The collinear parton distribution functions are then multiplied by the Gaussian function $g(k_{T})$,
\begin{eqnarray}
g(k_T) &=& \frac{1}{\pi \langle k_T^2 \rangle} \exp (k_T^2/\langle k_T^2\rangle) \;,
\end{eqnarray}
assuming the $x$ and $k_T$ dependences completely factorize. The same Gaussian smearing is applied in Refs.~\cite{Nelson:2012bc,Ma:2016exq,Mangano:1991jk}. Note that in the traditional CEM, the lower invariant mass threshold for all charmonium states is set to the production threshold, which makes the kinematic distributions of the charmonium states identical except for the choice of $F_{\mathcal{Q}}$. The distinction between the $J/\psi$ and $c\bar{c}$ momenta also helps describe the $p_T$ distributions at high $p_T$.

We consider diagrams with the projection operators, $\slashed{\epsilon}^*_\psi(J_z) (\slashed{p}_\psi+m_\psi)/(2m_\psi)$, applied to the $c\bar{c}$ \cite{Baranov:2002cf,Berger:1980ni} to calculate the partonic cross sections. We denote the momenta of $i$, $j$, $c$, $\bar{c}$, and $k$ in the partonic process $i+j\rightarrow c+\bar{c} +k$ are denoted as $k_1$, $k_2$, $p_c$, $p_{\bar{c}}$, and $k_3$, respectively, where $k$ is the emitted parton, with $\epsilon_1(S_{1z})$, $\epsilon_2(S_{2z})$, and $\epsilon_3(S_{3z})$ denoting the polarization of the light partons. When calculating the $2\rightarrow3$ cross section, we transformed the momenta of the charm quark ($p_c$) and the anticharm quark ($p_{\bar{c}}$) into the momentum of the proto-$J/\psi$ ($p_{\psi}$) and the relative momentum of the heavy quarks ($k_r$),
\begin{eqnarray}
p_c &=& \frac{1}{2}p_\psi+k_r \;, \\
p_{\bar{c}} &=& \frac{1}{2}p_\psi-k_r \;,
\end{eqnarray}
and denote the polarization vector of the proto-$J/\psi$ as $\epsilon_\psi(J_z)$, where $J_z$ is the spin projection onto the polarization axis. Instead of taking the limit $k_r\rightarrow0$, we note that, since the mass of the proto-$J/\psi$ is integrated from the physical mass of $J/\psi$ to the hadronic threshold, the relative momentum $k_r$ depends on the mass of the proto-$J/\psi$.

We factorized the amplitudes into a product of the color factors, $C$, and the colorless amplitude $\mathcal{A}$, so that $\mathcal{M} = C \mathcal{A}$. The color factors, $C$, in the squared amplitudes are calculated separately by summing over all colors and averaging over the initial-state colors.

There are 16 diagrams for the $gg \rightarrow c\bar{c} g$ process.  The factorized amplitudes, $\mathcal{A}$, arranged by the number of three-gluon vertices, are
\begin{widetext}
\begin{eqnarray}
\mathcal{A}_{gg0} &=& ig_s^3 \operatorname{tr} \Bigg[\slashed{\epsilon}_2(\slashed{p}_c-\slashed{k}_2+m_c)\slashed{\epsilon}_3^*(-\slashed{p}_{\bar{c}}+\slashed{k}_1+m_c)\slashed{\epsilon}_1 \slashed{\epsilon}^*_\psi(J_z) \frac{\slashed{p}_\psi+m_\psi}{2m_\psi} \Bigg] \frac{1}{2p_c\cdot k_2}\frac{1}{2p_{\bar{c}}\cdot k_1} \nonumber \\
&+& {\rm five~diagrams~with~no~three~\textendash~gluon~vertices} \;, \\
\mathcal{A}_{gg1} &=& ig_s^3 \operatorname{tr}\Bigg[\slashed{\epsilon}_3^*(\slashed{p}_c+\slashed{k}_3+m_c)\gamma^\mu \slashed{\epsilon}^*_\psi(J_z) \frac{\slashed{p}_\psi+m_\psi}{2m_\psi}\Bigg] [(-k_1-2k_2)\cdot \epsilon_1 \epsilon_{2\mu} \nonumber \\
&+& (\epsilon_1\cdot\epsilon_2) (-k_1+k_2)_\mu + (2k_1+k_2)\cdot\epsilon_2 \epsilon_{1\mu})] \frac{1}{2p_c\cdot k_3} \frac{1}{(k_1+k_2)^2} \nonumber \\
&+& {\rm five~diagrams~with~one~three~\textendash~gluon~vertex} \;, \\
\mathcal{A}_{gg2} &=& ig_s^3 \operatorname{tr} \Bigg[\gamma^\nu \slashed{\epsilon}^*_\psi(J_z) \frac{\slashed{p}_\psi+m_\psi}{2m_\psi}\Bigg][(-k_1-2k_2)\cdot \epsilon_1 \epsilon_2^\mu + (\epsilon_1\cdot\epsilon_2) (-k_1+k_2)^\mu \nonumber \\
&+& (2k_1+k_2)\cdot\epsilon_2 \epsilon_1^\mu)] \nonumber \\
&\times& [(-k_1-k_2-p_{\psi})\cdot \epsilon^*_3 g_{\mu\nu} + \epsilon_{3\mu}^* (k_3+k_1+k_2)_\nu+(p_\psi-k_3)_\mu \epsilon_{3\nu}^*]\frac{1}{(k_1+k_2)^2}\frac{1}{m_\psi^2} \nonumber \\
&+& {\rm two~diagrams~with~two~three~\textendash~gluon~vertices} \;.
\end{eqnarray}

The diagram with a four-gluon vertex is factorized according to
\begin{eqnarray}
\mathcal{M}_{gg4} &=& C_{gg4,1}\mathcal{A}_{gg4,1} + C_{gg4,2}\mathcal{A}_{gg4,2} + C_{gg4,3}\mathcal{A}_{gg4,3} \;,
\end{eqnarray}
with
\begin{eqnarray}
\mathcal{A}_{gg4,1} &=& ig_s^3 \operatorname{tr} \Bigg[\gamma^\nu \slashed{\epsilon}^*_\psi(J_z) \frac{\slashed{p}_\psi+m_\psi}{2m_\psi}\Bigg](g_{\alpha\gamma}g_{\nu\beta} - g_{\alpha\beta}g_{\nu\gamma})\frac{1}{m_\psi^2} \epsilon^\alpha \epsilon_2^\beta \epsilon_3^{*\gamma} \;, \\
\mathcal{A}_{gg4,2} &=& ig_s^3 \operatorname{tr} \Bigg[\gamma^\nu \slashed{\epsilon}^*_\psi(J_z) \frac{\slashed{p}_\psi+m_\psi}{2m_\psi}\Bigg](g_{\alpha\nu}g_{\gamma\beta} - g_{\alpha\beta}g_{\gamma\nu})\frac{1}{m_\psi^2} \epsilon^\alpha \epsilon_2^\beta \epsilon_3^{*\gamma} \;, \\
\mathcal{A}_{gg4,3} &=& ig_s^3 \operatorname{tr} \Bigg[\gamma^\nu \slashed{\epsilon}^*_\psi(J_z) \frac{\slashed{p}_\psi+m_\psi}{2m_\psi} \Bigg](g_{\alpha\gamma}g_{\nu\beta} - g_{\alpha\nu}g_{\gamma\beta})\frac{1}{m_\psi^2} \epsilon^\alpha \epsilon_2^\beta \epsilon_3^{*\gamma} \;.
\end{eqnarray}

There are five $gq \rightarrow c\bar{c} q$ diagrams, which, written in terms of Dirac spinors, are
\begin{eqnarray}
\mathcal{A}_{gq,1} &=& -ig_s^3 \operatorname{tr} \Bigg[\gamma^\nu \slashed{\epsilon}^*_\psi(J_z) \frac{\slashed{p}_\psi+m_\psi}{2m_\psi}\Bigg] [\bar{u}(k_3) \gamma_\nu (\slashed{k}_1+\slashed{k}_2) \slashed{\epsilon}_1 u(k_2)] \frac{1}{m_\psi^2}\frac{1}{2k_1\cdot k_2} \;, \\
\mathcal{A}_{gq,2} &=& -ig_s^3 \operatorname{tr} \Bigg[\slashed{\epsilon}_1(-\slashed{p}_c-\slashed{k}_1+m_c) \gamma^\nu \slashed{\epsilon}^*_\psi(J_z) \frac{\slashed{p}_\psi+m_\psi}{2m_\psi}\Bigg] [\bar{u}(k_3) \gamma_\nu u(k_2)]  \nonumber \\
&\times&\frac{1}{-2p_c\cdot k_1} \frac{1}{(-k_3+k_2)^2} \;, \\
\mathcal{A}_{gq,3} &=& -ig_s^3 \operatorname{tr} \Bigg[\gamma^\nu(\slashed{p}_{\bar{c}}+\slashed{k}_1+m_c) \slashed{\epsilon}_1 \slashed{\epsilon}^*_\psi(J_z) \frac{\slashed{p}_\psi+m_\psi}{2m_\psi}\Bigg] [\bar{u}(k_3) \gamma_\nu u(k_2)]  \nonumber \\
&\times&\frac{1}{-2p_{\bar{c}}\cdot k_1} \frac{1}{(-k_3+k_2)^2} \;, \\
\mathcal{A}_{gq,4} &=& -ig_s^3 \operatorname{tr} \Bigg[\gamma^\nu \slashed{\epsilon}^*_\psi(J_z) \frac{\slashed{p}_\psi+m_\psi}{2m_\psi}\Bigg] [(k_3-k_2-p_\psi)\cdot \epsilon_1 g_{\mu\nu} + \epsilon_{1\mu}(-k_1-k_3+k_2)_\nu  \nonumber \\
&+&(p_\psi+k_1)_\mu \epsilon_{1\nu} ]  [\bar{u}(k_3) \gamma^\mu u(k_2)] \frac{1}{m_\psi^2} \frac{1}{(-k_3+k_2)^2}  \;, \\
\mathcal{A}_{gq,5} &=& -ig_s^3 \operatorname{tr} \Bigg[\gamma^\nu \slashed{\epsilon}^*_\psi(J_z) \frac{\slashed{p}_\psi+m_\psi}{2m_\psi}\Bigg] [\bar{u}(k_3)\slashed{\epsilon}_1 (\slashed{k}_3-\slashed{k}_1)\gamma_\nu u(k_2)] \frac{1}{m_\psi^2}\frac{1}{-2k_1\cdot k_2} \;.
\end{eqnarray}

There are five $g\bar{q} \rightarrow c\bar{c} \bar{q}$ diagrams, obtained by replacing the spinors in the above five $gq \rightarrow c\bar{c} q$ diagrams:
\begin{eqnarray}
\mathcal{A}_{g\bar{q},1} &=& -ig_s^3 \operatorname{tr} \Bigg[\gamma^\nu \slashed{\epsilon}^*_\psi(J_z) \frac{\slashed{p}_\psi+m_\psi}{2m_\psi}\Bigg] [\bar{v}(k_2) \slashed{\epsilon}_1 (-\slashed{k}_2-\slashed{k}_1) \gamma_\nu v(k_3)] \frac{1}{m_\psi^2}\frac{1}{2k_1\cdot k_2} \;, \\
\mathcal{A}_{g\bar{q},2} &=& -ig_s^3 \operatorname{tr} \Bigg[\slashed{\epsilon}_1(\slashed{p}_c-\slashed{k}_1+m_c)\gamma^\nu\slashed{\epsilon}^*_\psi(J_z) \frac{\slashed{p}_\psi+m_\psi}{2m_\psi}\Bigg][\bar{v}(k_2)\gamma_\nu v(k_3)] \nonumber \\
&\times& \frac{1}{-2p_c\cdot k_1}\frac{1}{(-k_3+k_2)^2} \;, \\
\mathcal{A}_{g\bar{q},3} &=& -ig_s^3 \operatorname{tr} \Bigg[\gamma^\nu (-\slashed{p}_{\bar{c}}+\slashed{k}_1+m_c)\slashed{\epsilon}_1\slashed{\epsilon}^*_\psi(J_z) \frac{\slashed{p}_\psi+m_\psi}{2m_\psi}\Bigg] [\bar{v}(k_2)\gamma_\nu v(k_3)] \nonumber \\
&\times& \frac{1}{-2p_{\bar{c}}\cdot k_1}\frac{1}{(-k_3+k_2)^2} \;, \\
\mathcal{A}_{g\bar{q},4} &=& -ig_s^3 \operatorname{tr} \Bigg[\gamma^\nu \slashed{\epsilon}_1\slashed{\epsilon}^*_\psi(J_z) \frac{\slashed{p}_\psi+m_\psi}{2m_\psi}\Bigg] [(k_3-k_2-p_\psi)\cdot \epsilon_1 g_{\mu\nu} \nonumber \\
&+& \epsilon_{1\mu}(-k_1-k_3+k_2)_\nu+(p_\psi+k_1)_\mu\epsilon_{1\nu}] [\bar{v}(k_2)\gamma^\mu v(k_3)]\frac{1}{m_\psi^2}\frac{1}{(-k_3+k_2)^2} \;, \\
\mathcal{A}_{g\bar{q},5} &=& -ig_s^3 \operatorname{tr} \Bigg[\gamma^\nu\slashed{\epsilon}_1\slashed{\epsilon}^*_\psi(J_z) \frac{\slashed{p}_\psi+m_\psi}{2m_\psi}\Bigg] [\bar{v}(k_2)\gamma_\nu(-\slashed{k}_3+\slashed{k}_1)\slashed{\epsilon}_1v(k_3)] \nonumber \\
&\times& \frac{1}{m_\psi^2}\frac{1}{-2k_3\cdot k_1} \;.
\end{eqnarray}
Finally, five diagrams contribute to $q\bar{q} \rightarrow c\bar{c} g$,
\begin{eqnarray}
\mathcal{A}_{q\bar{q},1} &=& -ig_s^3 \operatorname{tr} \Bigg[\gamma^\nu \slashed{\epsilon}^*_\psi(J_z) \frac{\slashed{p}_\psi+m_\psi}{2m_\psi}\Bigg] [(-k_1-k_2-p_\psi)\cdot \epsilon_3^* g_{\mu\nu} \nonumber \\
&+& \epsilon_{3\mu}^* (k_3+k_1+k_2)_\nu +(p_\psi-k_3)_\mu \epsilon_{3\nu}^*] [\bar{v}(k_2)\gamma^\mu u(k_1)]\frac{1}{m_\psi^2}\frac{1}{(k_1+k_2)^2} \;, \\
\mathcal{A}_{q\bar{q},2} &=& -ig_s^3 \operatorname{tr} \Bigg[\slashed{\epsilon}_3^* (\slashed{p}_c+\slashed{k}_3+m_c)\gamma^\nu\slashed{\epsilon}^*_\psi(J_z) \frac{\slashed{p}_\psi+m_\psi}{2m_\psi}\Bigg] [\bar{v}(k_2)\gamma_\nu u(k_1)] \nonumber \\
&\times& \frac{1}{2p_c\cdot k_3}\frac{1}{(k_1+k_2)^2} \;, \\
\mathcal{A}_{q\bar{q},3} &=& -ig_s^3 \operatorname{tr} \Bigg[\gamma^\nu(-\slashed{p}_{\bar{c}}-\slashed{k}_3+m_c)\slashed{\epsilon}_3^* \slashed{\epsilon}^*_\psi(J_z) \frac{\slashed{p}_\psi+m_\psi}{2m_\psi}\Bigg] [\bar{v}(k_2)\gamma_\nu u(k_1)] \nonumber \\
&\times& \frac{1}{2 p_{\bar{c}}\cdot k_3}\frac{1}{(k_1+k_2)^2} \;, \\
\mathcal{A}_{q\bar{q},4} &=& -ig_s^3 \operatorname{tr} \Bigg[\gamma^\nu\slashed{\epsilon}^*_\psi(J_z) \frac{\slashed{p}_\psi+m_\psi}{2m_\psi}\Bigg] [\bar{v}(k_2)\gamma_\nu(-\slashed{k}_3+\slashed{k}_1)\slashed{\epsilon}_3^*u(k_1)] \nonumber \\
&\times& \frac{1}{m_\psi^2}\frac{1}{-2k_1\cdot k_3} \nonumber \\
\mathcal{A}_{q\bar{q},5} &=& -ig_s^3 \operatorname{tr} \Bigg[\gamma^\nu\slashed{\epsilon}^*_\psi(J_z) \frac{\slashed{p}_\psi+m_\psi}{2m_\psi}\Bigg] [\bar{v}(k_2)\slashed{\epsilon}_3^*(-\slashed{k}_2+\slashed{k}_3)\gamma_\nu v(k_1)] \nonumber \\
&\times& \frac{1}{m_\psi^2}\frac{1}{-2k_2\cdot k_3} \;.
\end{eqnarray}
\end{widetext}
We assume that the angular momentum of the proto-$J/\psi$ is unchanged by the transition from the parton level to the hadron level. We then convolute the partonic cross sections with the CT14 PDFs \cite{Dulat:2015mca} in the domain where $p_\psi \cdot k =0$. We restrict the partonic cross section calculations within the perturbative domain by introducing a regularization parameter such that all propagators are at a minimum distance of $Q_{\rm reg}^2=M^2$ from their poles, as employed in Ref.~\cite{Baranov:2002cf}. We take the factorization and renormalizaton scales to be $\mu_F/m_T = 2.1^{+2.55}_{-0.85}$ and $\mu_F/m_T = 1.6^{+0.11}_{-0.12}$ respectively, where $m_T$ is the transverse mass of the charm quark produced ($m_T = \sqrt{m_c^2+p_T^2}$, where $p_T^2 = 0.5\sqrt{p_{Tc}^2+p_{T\bar{c}}^2}$). We also vary the charm quark mass around $1.27\pm 0.09$~GeV. These variations were determined in Ref.~\cite{Nelson:2012bc}, in which the uncertainties on the total charm cross section were considered.
\label{ch6-section2}

\section{Polarized production of direct $J/\psi$}
\label{section3}
We factor the polarization vector, $\epsilon_\psi(J_z)$, from the unsquared amplitudes for all sub processes, giving us the form
\begin{eqnarray}
\mathcal{M}_n &=& \epsilon_\psi^\mu(J_z) \mathcal{M}_{n,\mu}
\end{eqnarray}
for each subprocess denoted by the initial states, $n=gg,gq,g\bar{q},q\bar{q}$. The polarization vectors for $J_z=0$, $\pm1$ in the rest frame of the proto-$J/\psi$ are
\begin{eqnarray}
\epsilon_\psi(0)^\mu &=& (1,0,0,0) \;, \\
\epsilon_\psi(\pm1)^\mu &=& \mp\frac{1}{\sqrt{2}}(0,1,\pm i,0) \;,
\end{eqnarray}
using the convention that the fourth component is the $z$ component. While the unpolarized cross section does not depend on the choice of $z$ axis, the polarized cross sections does depend on the orientation of the $z$ axis. In this calculation, the $y$ axis is chosen to be the normal vector of the plane formed by the two beams with momenta $\vec{P}_1$ and $\vec{P}_2$,
\begin{eqnarray}
\hat{y} &=& \frac{-\vec{P}_1\times\vec{P}_2}{|\vec{P}_1\times\vec{P}_2|} \;.
\end{eqnarray}
In the helicity frame, the $z_{HX}$ axis is the flight direction of the $c\bar{c}$ pair in the center of mass of the colliding beams. In the Collins-Soper frame \cite{Collins:1977iv}, the $z_{CS}$ axis is the angle bisector between one beam and the opposite direction of the other beam. The $x$ axis is then determined by the right-handed convention.

We compute the polarized cross section matrix element, $\mathcal{M}_n$, in the rest frame of the $c\bar{c}$ pair by first taking the product of the unsquared amplitude with polarization vector of $J_z=i_z$ and the unsquared amplitude with polarization vector of $J_z=j_z$ in each subprocess ($n$), then adding them, and finally calculating the components of the polarized cross section matrix according to Eq.~(\ref{ch6-icem-cross-section}),
\begin{eqnarray}
\sigma_{i_z,j_z} &=& \int \sum_{n} (\epsilon_\psi^\mu(i_z) \mathcal{M}_{n,\mu})(\epsilon_\psi^{\nu}(j_z) \mathcal{M}_{n,\nu})^* \;,
\end{eqnarray}
where $i_z,j_z=\{-1,0,+1\}$ and the integral is over all variables explicitly shown in Eq.~(\ref{ch6-icem-cross-section}) as well as the Lorentz-invariant phase space in $2\rightarrow3$ scatterings. The unpolarized cross section is the trace of the polarized cross section matrix
\begin{eqnarray}
\sigma_{\rm unpol} &=& \sum_{i_z} \sigma_{i_z,i_z} = \sigma_{-1,-1} + \sigma_{0,0} + \sigma_{+1,+1} \;.
\end{eqnarray}

The polarization parameters are calculated using the matrix elements. The polar anisotropy ($\lambda_{\vartheta}$), the azimuthal anisotropy ($\lambda_\varphi$), and polar-azimuthal correlation ($\lambda_{\vartheta\varphi}$) are given by \cite{Faccioli:2010kd}
\begin{eqnarray}
\lambda_{\vartheta} &=& \frac{\sigma_{+1,+1}-\sigma_{0,0}}{\sigma_{+1,+1}+\sigma_{0,0}} \; \label{lambda_theta_eqn} ,\\
\lambda_{\varphi} &=& \frac{\operatorname{Re}[\sigma_{+1,-1}]}{\sigma_{+1,+1}+\sigma_{0,0}} \;, \\
\lambda_{\vartheta\varphi} &=& \frac{\operatorname{Re}[\sigma_{+1,0}-\sigma_{-1,0}]}{\sqrt{2}(\sigma_{+1,+1}+\sigma_{0,0})} \label{lambda_theta_phi_eqn}\;.
\end{eqnarray}
These parameters depend on the frame (helicity or Collins-Soper) in which they are calculated and measured. Since the angular distribution itself is rotationally invariant, there are ways to construct invariant polarization parameters from Eqs.~(\ref{lambda_theta_eqn})–(\ref{lambda_theta_phi_eqn}). One of the combinations to form a frame-invariant polarization parameter ($\tilde{\lambda}$) is \cite{Faccioli:2010kd}
\begin{eqnarray}
\tilde{\lambda} &=& \frac{\lambda_\vartheta+3\lambda_\varphi}{1 - \lambda_\varphi} \;.
\end{eqnarray}
The choice of $\tilde{\lambda}$ is the same as the polar anisotropy parameter ($\lambda_\vartheta$) in a frame where the distribution is azimuthally isotropic ($\lambda_{\varphi}=0$). We can remove the frame-induced kinematic dependencies when comparing theoretical predictions to data by also considering the frame-invariant polarization parameter, $\tilde{\lambda}$.

\section{Results}
\label{section4}
We first show how our approach describes the transverse-momentum-distributions of $J/\psi$ compared to ALICE \cite{Aamodt:2011gj} and ATLAS \cite{Aad:2015duc} measurements at $\sqrt{s}=7$~TeV and compare our results with previous calculations in the ICEM. We then discuss the transverse momentum and rapidity dependences of the frame-dependent polarization parameters $\lambda_\vartheta$, $\lambda_{\varphi}$, and $\lambda_{\vartheta\varphi}$ as well as the frame-invariant polarization parameter $\tilde{\lambda}$ compared to the data measured by the LHCb Collaboration \cite{Aaij:2013nlm}, the ALICE Collaboration \cite{Abelev:2011md}, and the CMS Collaboration \cite{Chatrchyan:2013cla}. In our calculations, we consider theoretical uncertainties by varying the charm quark mass, the renormalization scale, and the factorization scale as discussed in Sec.~\ref{ch6-section2}. We also estimate the uncertainty on the fit parameter, $\delta F_{J\psi}$. The total uncertainty band is then constructed by adding the mass and scale uncertainties in quadrature and varying $F_{J/\psi}$ from $F_{J/\psi}-\delta F_{J\psi}$ to $F_{J/\psi}+\delta F_{J\psi}$.

\subsection{Unpolarized $J/\psi$ $p_T$ distribution}

\begin{figure}[b!]
\centering
\includegraphics[width=\columnwidth]{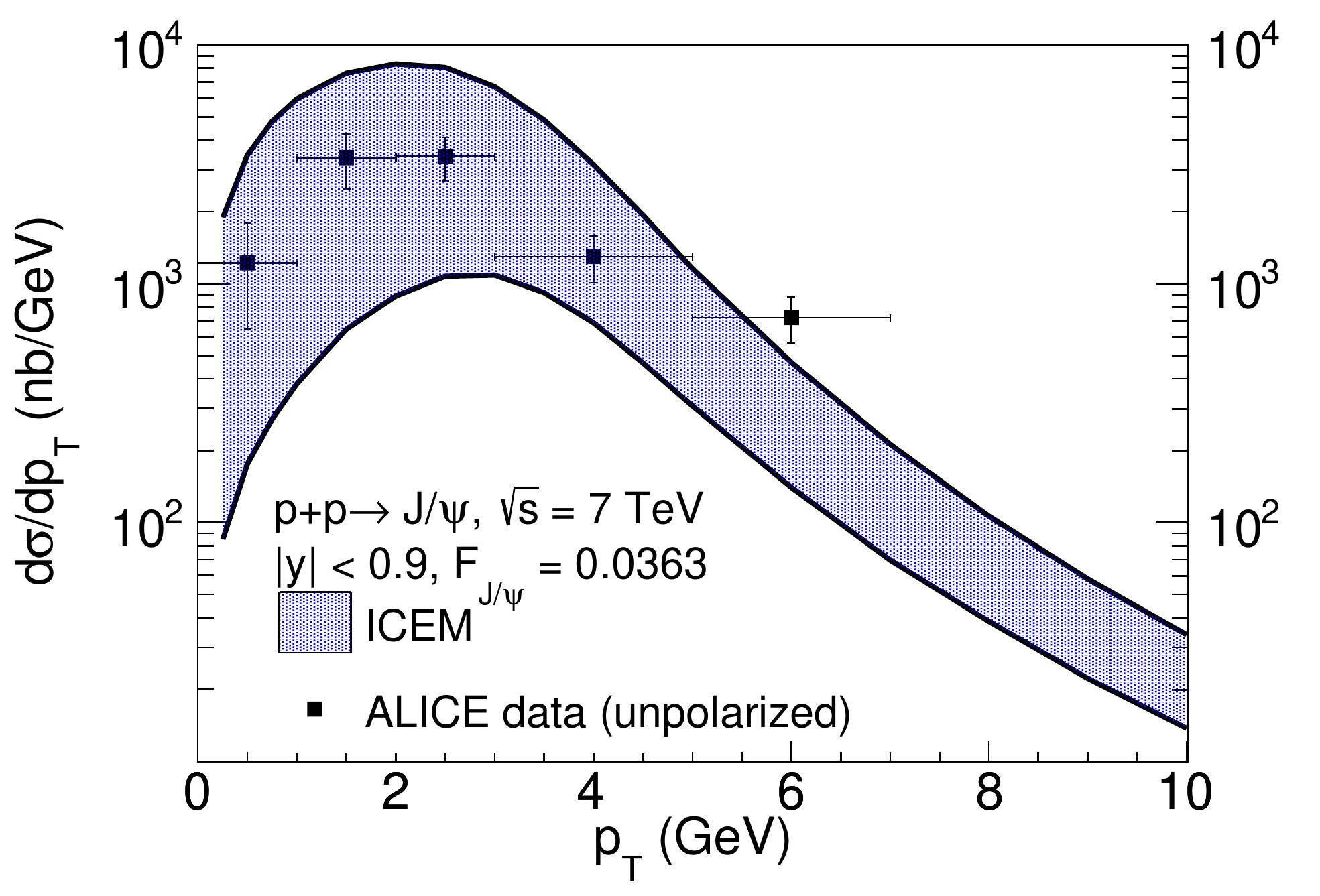}
\caption{The $p_T$ dependence of inclusive $J/\psi$ production at $\sqrt{s} = 7$~TeV with $|y|<0.9$ in the ICEM. The combined mass, renormalization scale, factorization scale, and $F_{J/\psi}$ uncertainties are shown in the band and compared to the ALICE data \cite{Aamodt:2011gj}. The ALICE data are measured while assuming $J/\psi$ production is unpolarized.} \label{ch6-ALICE-pt-dist}
\end{figure}

We calculate the $p_T$ distribution of direct $J/\psi$ production at $\sqrt{s}=7$~TeV with $|y|<0.9$. We assume direct production is a constant fraction, $0.62$, of the inclusive production \cite{Digal:2001ue} to obtain the inclusive $J/\psi$ $p_T$ distribution. We compare our ICEM inclusive $J/\psi$ $p_T$ distribution with the data measured by the ALICE Collaboration \cite{Aamodt:2011gj}. The comparison is presented in Fig.~\ref{ch6-ALICE-pt-dist}. By comparing the total cross section for $p_T<7$~GeV, we find $F_{J/\psi} = 0.0363$, consistent with previous CEM \cite{Nelson:2012bc} and ICEM calculations \cite{Cheung:2018tvq}. We add the statistical and systematic uncertainties of the ALICE total cross section in quadrature. This results in a proportional uncertainty of 17\% on the total cross section. Using the same proportional uncertainty, we estimate the uncertainty on $F_{J/\psi}$, $\delta F_{J\psi}$ to be $\pm0.0062$. We note that since this calculation is done using collinear factorization, the variation in the factorization scale does not result in a large uncertainty band as seen in our previous calculation using the $k_T$-factorization approach \cite{Cheung:2018tvq}. Overall, we have good agreement with the data over the $p_T$ range measured.

\begin{figure}[t!]
\centering
\includegraphics[width=\columnwidth]{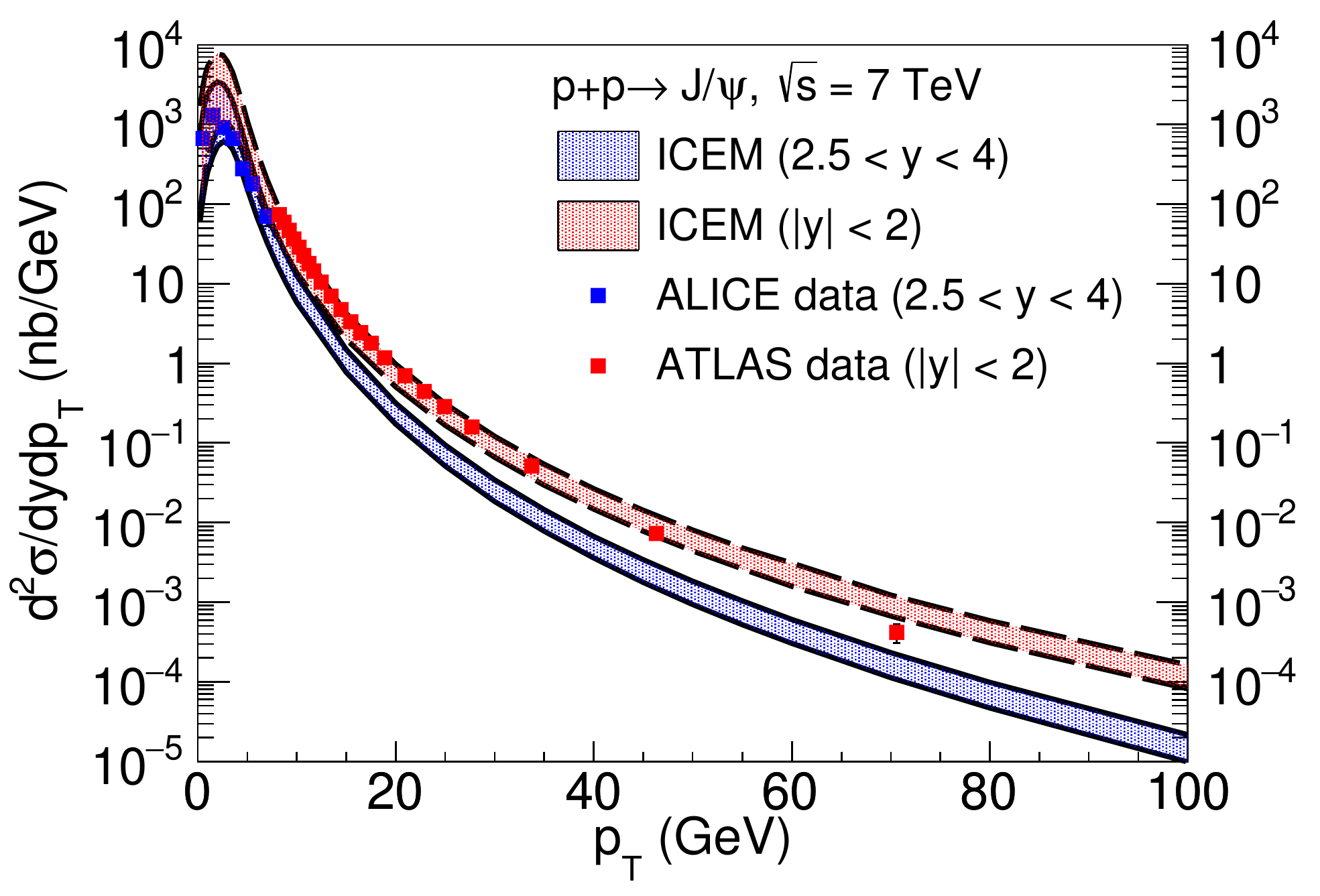}
\caption{The $p_T$ dependence of inclusive $J/\psi$ production at $\sqrt{s} = 7$~TeV with $2.5<y<4$ (blue region) and $|y| < 2$ in the ICEM (red region). The combined mass, renormalization scale, factorization scale, and $F_{J/\psi}$ uncertainties are shown in the band. They are compared to the ALICE data \cite{Aamodt:2011gj} (blue squares) and the ATLAS data \cite{Aad:2015duc} (red squares) respectively.} \label{p_T-dists-compare}
\end{figure}

We compare the ICEM inclusive $J/\psi$ $p_T$ distributions at $\sqrt{s}=7$~TeV in the forward rapidity region covered by the ALICE detector, $2.5<y<4$ \cite{Aamodt:2011gj}, and in the central rapidity region covered by the ATLAS detector, $|y|<2$ \cite{Aamodt:2011gj}. These comparisons are presented in Fig.~\ref{p_T-dists-compare}. Both ALICE and ATLAS data are shown assuming $J/\psi$ production is unpolarized. Since the detector acceptance depends on the polarization assumption, the ATLAS Collaboration reports a set of correction factors for each polarization assumptions. These factors ranges from 0.7 to 1.7 for the lowest $p_T$ bin, and from 0.90 to 1.06 for the highest $p_T$ bin. Except for the highest $p_T$ bin reported by the ATLAS Collaboration, we find our ICEM results in good agreement with the data over the $p_T$ range measured in both kinematic regions using the same $F_{J/\psi}$ found by comparing to the ALICE central rapidity data.

\begin{figure}[b!]
\centering
\includegraphics[width=\columnwidth]{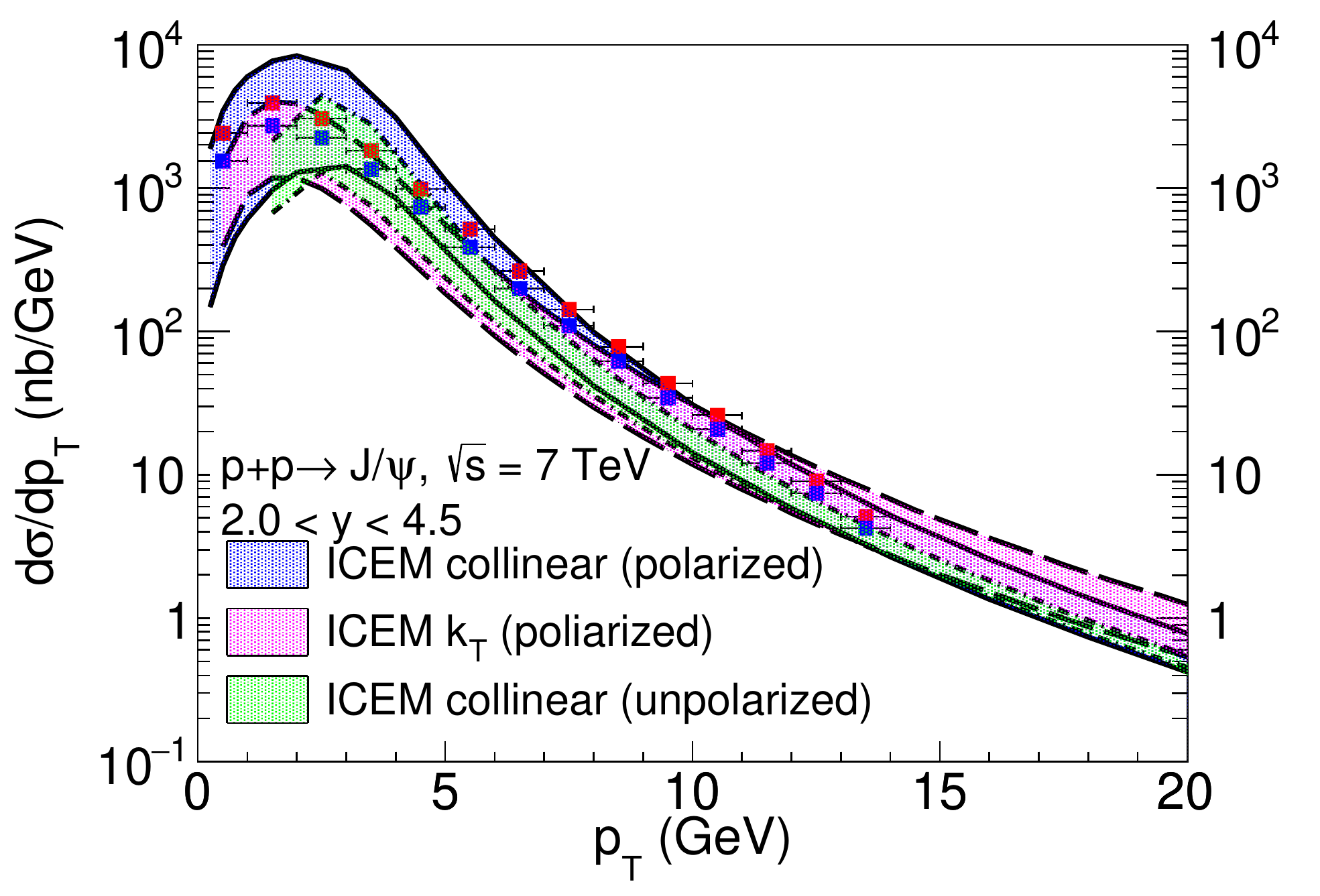}
\caption{The $p_T$ dependence of inclusive $J/\psi$ production at $\sqrt{s} = 7$~TeV in the polarized collinear ICEM (this calculation) (blue region), in the polarized ICEM using the $k_T$-factorization \cite{Cheung:2018tvq} (magenta region), in the unpolarized collinear ICEM \cite{Ma:2016exq} (green region). They are compared to the LHCb data \cite{Aaij:2011jh} assuming that the $J/\psi$ polarization is totally transverse, $\lambda_\vartheta = +1$ (red squares), and totally longitudinal, $\lambda_\vartheta = -1$ (blue squares). The LHCb data assuming $\lambda_\vartheta = 0$ lie between the red and blue points and are not shown.} \label{ch6-LHCb_1S_pt_CEM_compare}
\end{figure}

We compare the $p_T$ distribution of inclusive $J/\psi$ production at $\sqrt{s}=7$~TeV with $2<y<4.5$ in this calculation with that from previous calculations in the ICEM shown in Fig.~\ref{ch6-LHCb_1S_pt_CEM_compare}. We compare these calculations with the data measured by the LHCb Collaboration \cite{Aaij:2011jh}. Although this distribution is for unpolarized $J/\psi$, the calculation selects only $c\bar{c}$ with the same spin as the $J/\psi$. We thus refer to this calculation as polarized collinear ICEM. We compare this distribution with the polarized ICEM in the $k_T$-factorization approach \cite{Cheung:2018tvq} and the unpolarized ICEM in the collinear factorization approach \cite{Ma:2016exq}. The former also selects $c\bar{c}$ with the same spin as the $J/\psi$ but the latter is a spin-averaged calculation. The uncertainty band of the unpolarized collinear ICEM is constructed in the same way as the polarized collinear ICEM. The uncertainty band of the $k_T$-factorized ICEM is constructed by varying the renormalization scale in the interval $0.5<\mu_R/m_T<2$ and varying the charm mass in the interval $1.2<m_c<1.5$. We find our polarized collinear ICEM $p_T$ distribution behaves very similarly to the unpolarized collinear ICEM as $p_T$ increases and agrees with the data and other calculations in the ICEM in general.

\begin{figure*}
\centering
\begin{minipage}[ht]{0.68\columnwidth}
\centering
\includegraphics[width=\columnwidth]{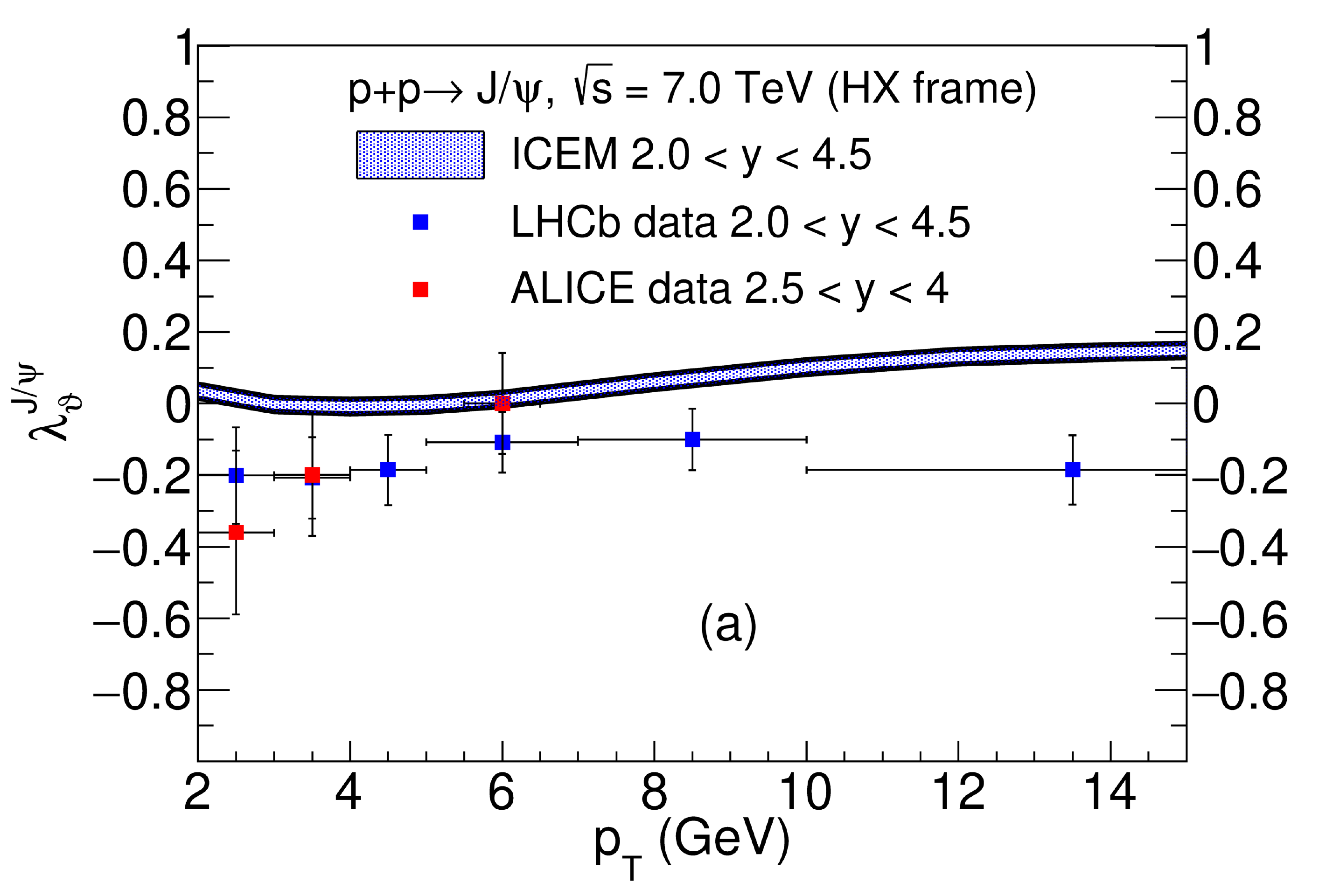}
\end{minipage}%
\begin{minipage}[ht]{0.68\columnwidth}
\centering
\includegraphics[width=\columnwidth]{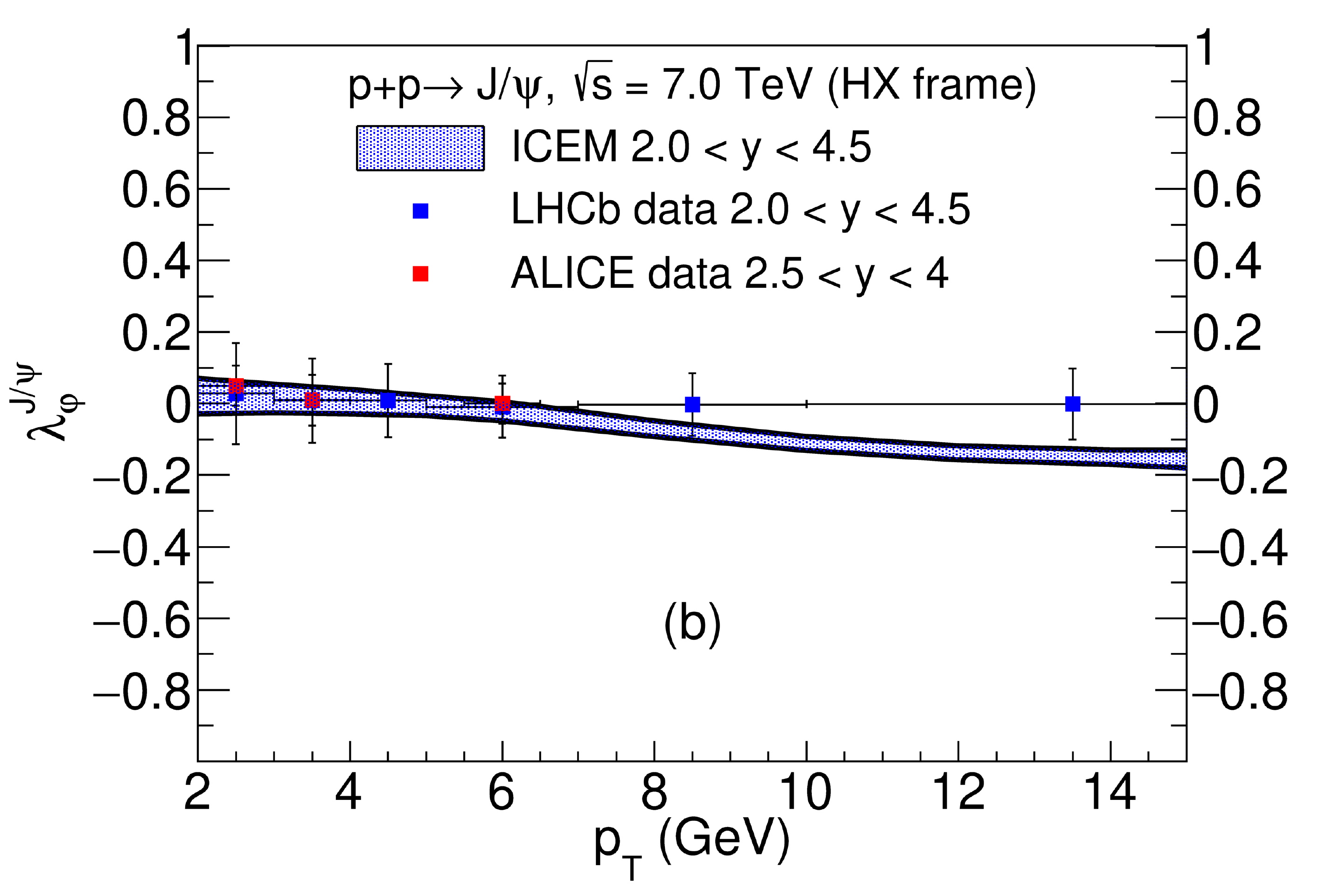}
\end{minipage}
\begin{minipage}[ht]{0.68\columnwidth}
\centering
\includegraphics[width=\columnwidth]{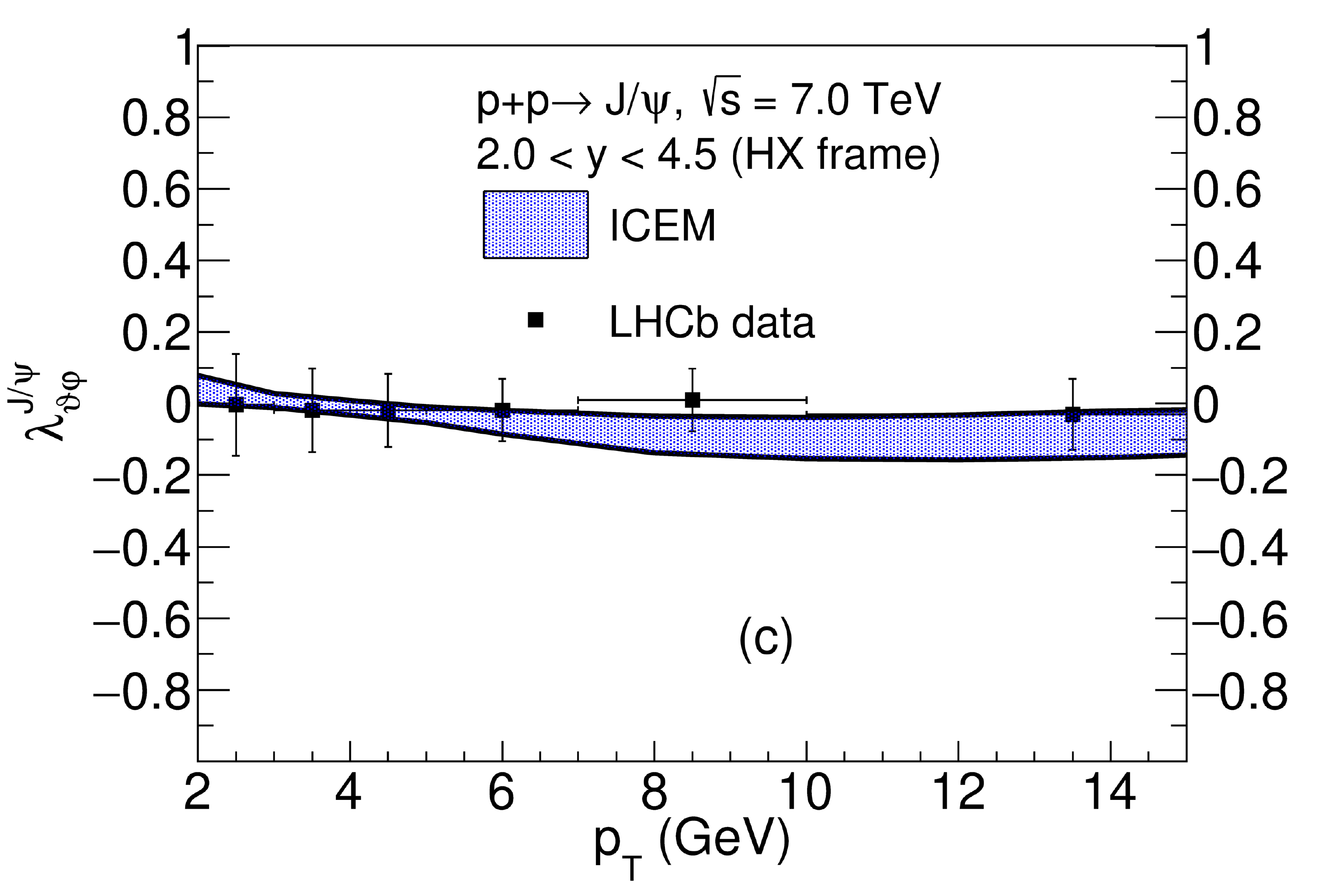}
\end{minipage}
\caption{(a) The polar anisotropy parameter ($\lambda_\vartheta$), (b) the azimuthal anisotropy parameter ($\lambda_\varphi$), and (c) the polar-azimuthal correlation parameter ($\lambda_{\vartheta\varphi}$) (c) in the helicity frame at $\sqrt{s} = 7$~TeV in the ICEM. The combined mass, renormalization scale, and factorization scale uncertainties are shown in the band and compared to the LHCb data \cite{Aaij:2013nlm} (blue) and the ALICE data \cite{Abelev:2011md} (red).} \label{frame-dependent-lambdas-hx}
\end{figure*}

\begin{figure*}
\centering
\begin{minipage}[ht]{0.68\columnwidth}
\centering
\includegraphics[width=\columnwidth]{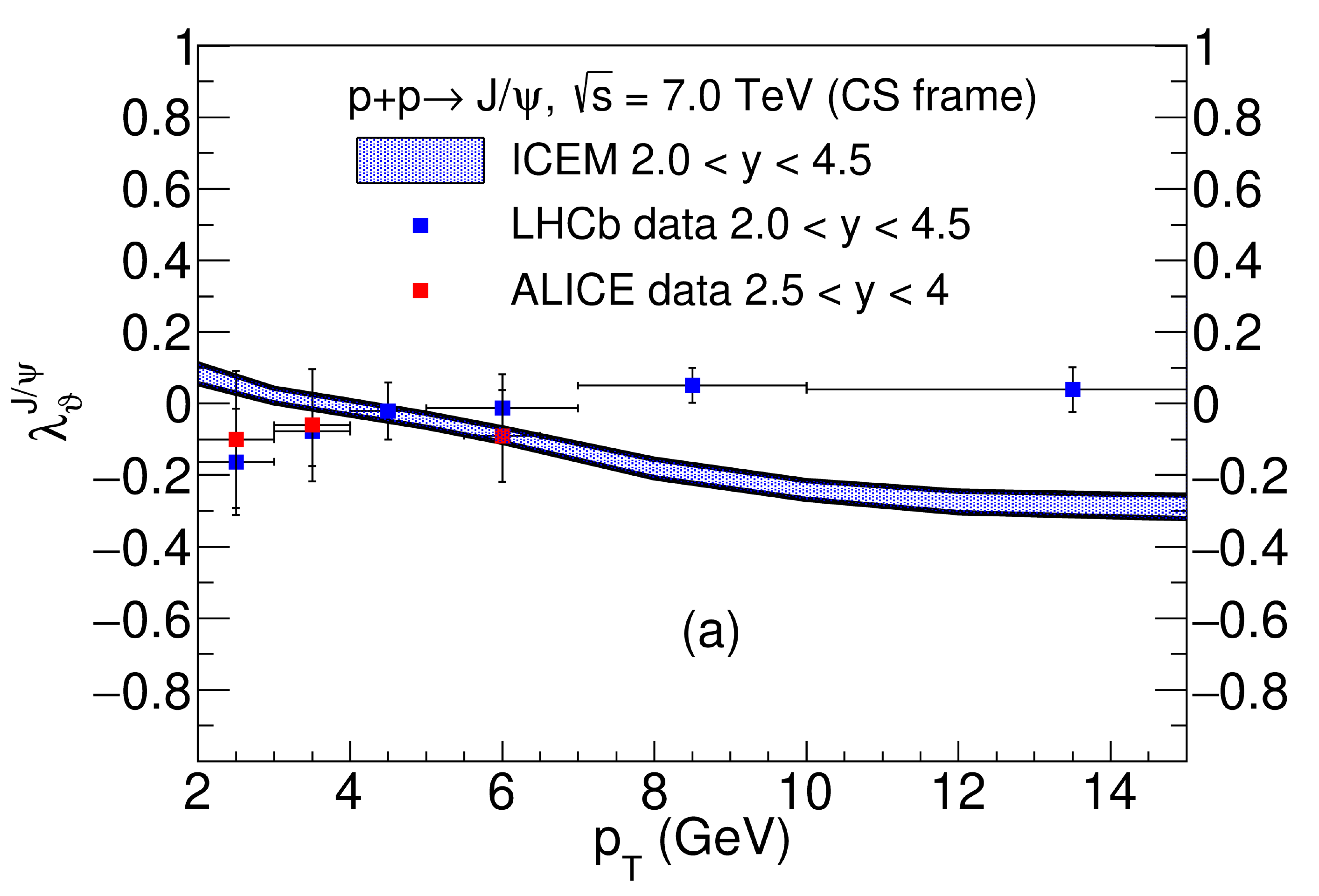}
\end{minipage}%
\begin{minipage}[ht]{0.68\columnwidth}
\centering
\includegraphics[width=\columnwidth]{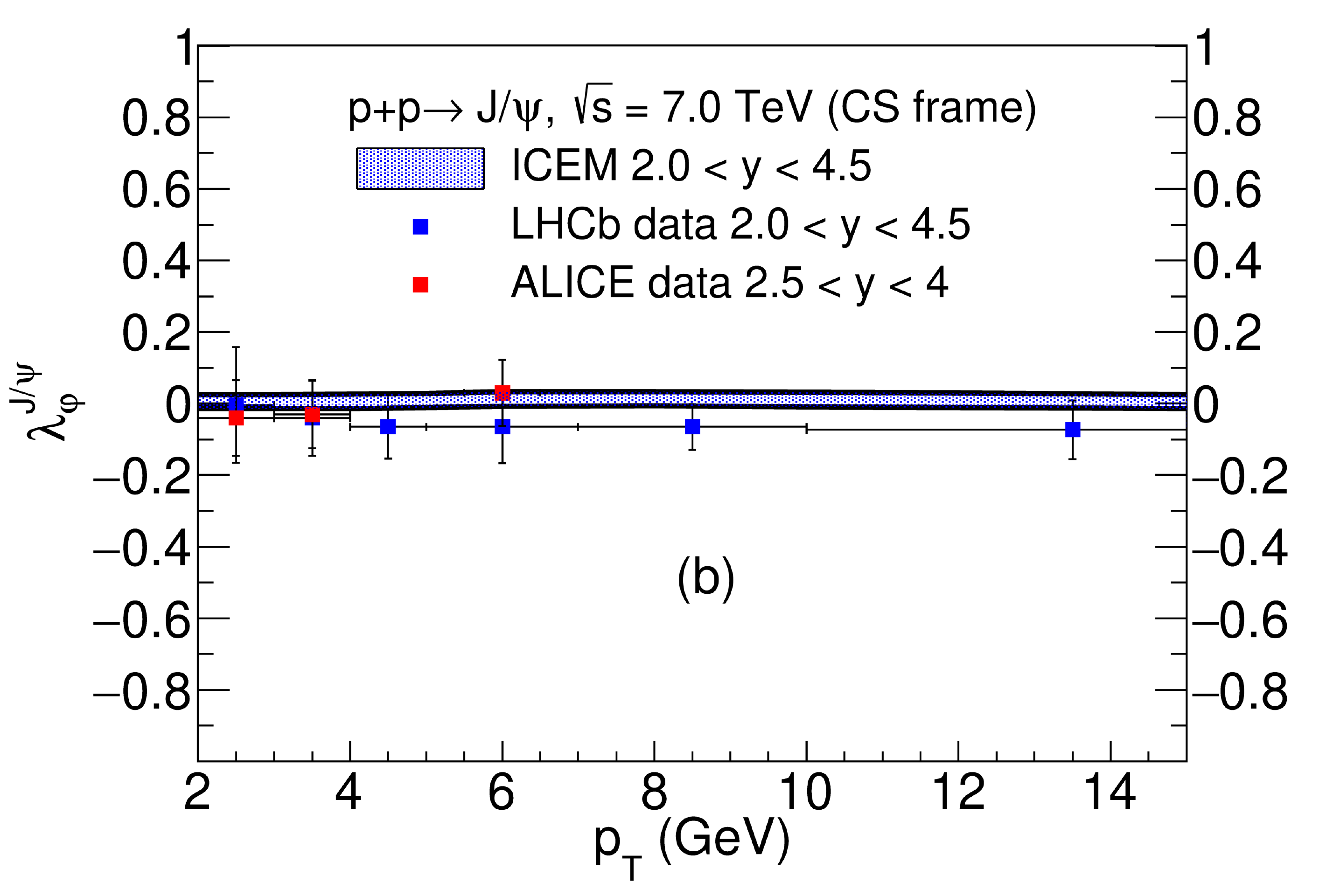}
\end{minipage}
\begin{minipage}[ht]{0.68\columnwidth}
\centering
\includegraphics[width=\columnwidth]{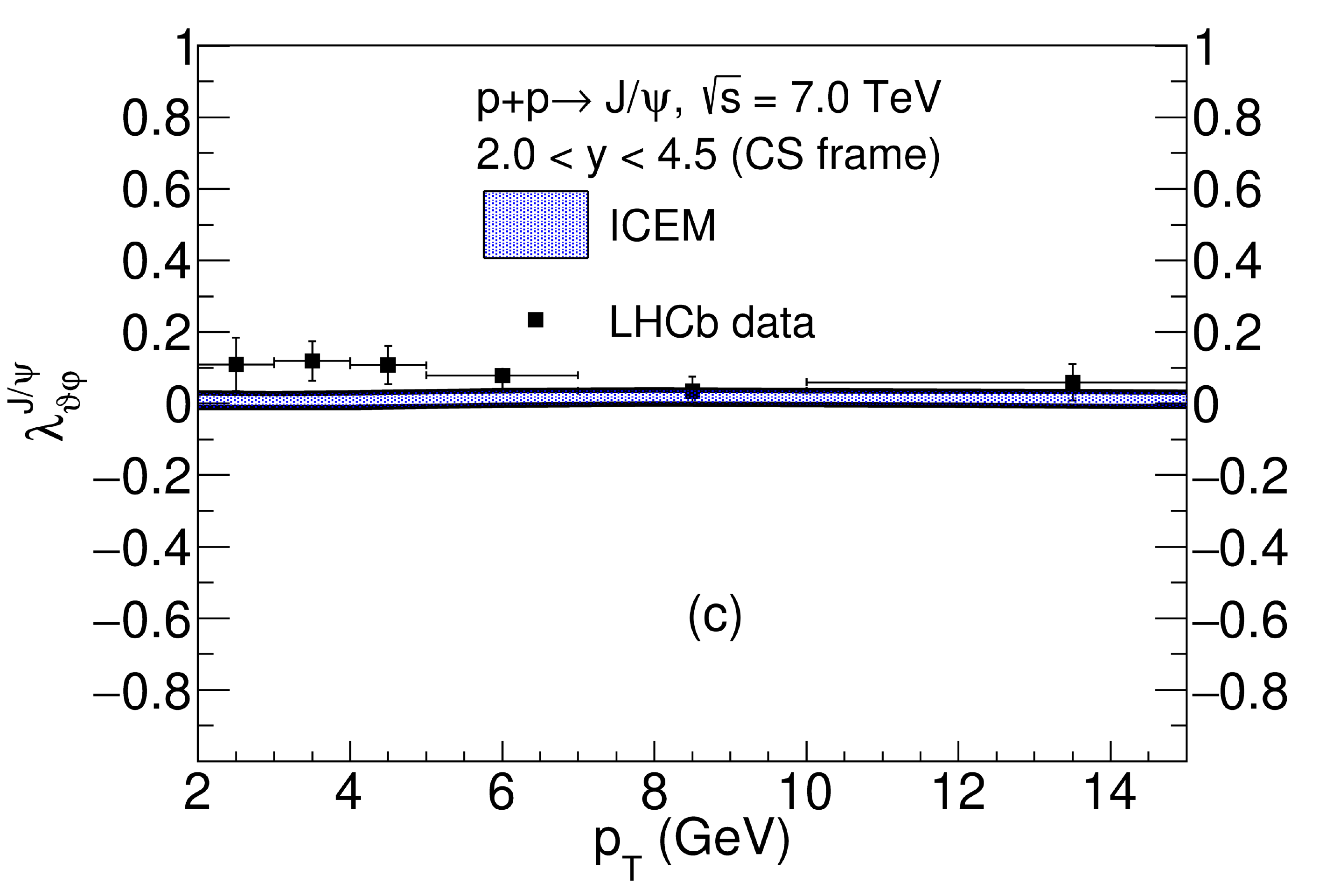}
\end{minipage}
\caption{(a) The polar anisotropy parameter ($\lambda_\vartheta$), (b) the azimuthal anisotropy parameter ($\lambda_\varphi$), and (c) the polar-azimuthal correlation parameter ($\lambda_{\vartheta\varphi}$) (c) in the Collins-Soper frame at $\sqrt{s} = 7$~TeV in the ICEM. The combined mass, renormalization scale, and factorization scale uncertainties are shown in the band and compared to the LHCb data \cite{Aaij:2013nlm} (blue) and the ALICE data \cite{Abelev:2011md} (red).} \label{frame-dependent-lambdas-cs}
\end{figure*}

\begin{figure*}
\centering
\begin{minipage}[ht]{0.68\columnwidth}
\centering
\includegraphics[width=\columnwidth]{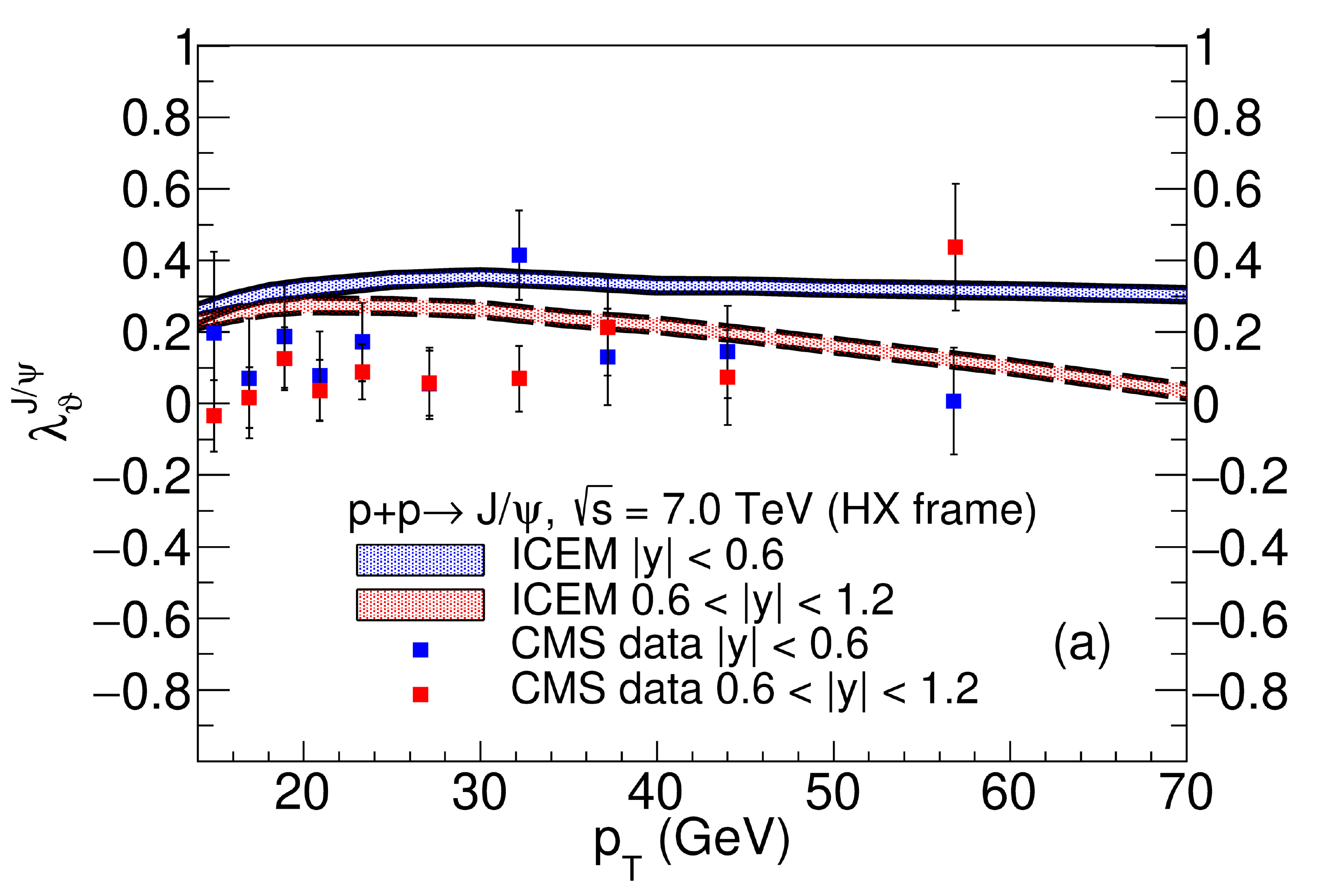}
\end{minipage}%
\begin{minipage}[ht]{0.68\columnwidth}
\centering
\includegraphics[width=\columnwidth]{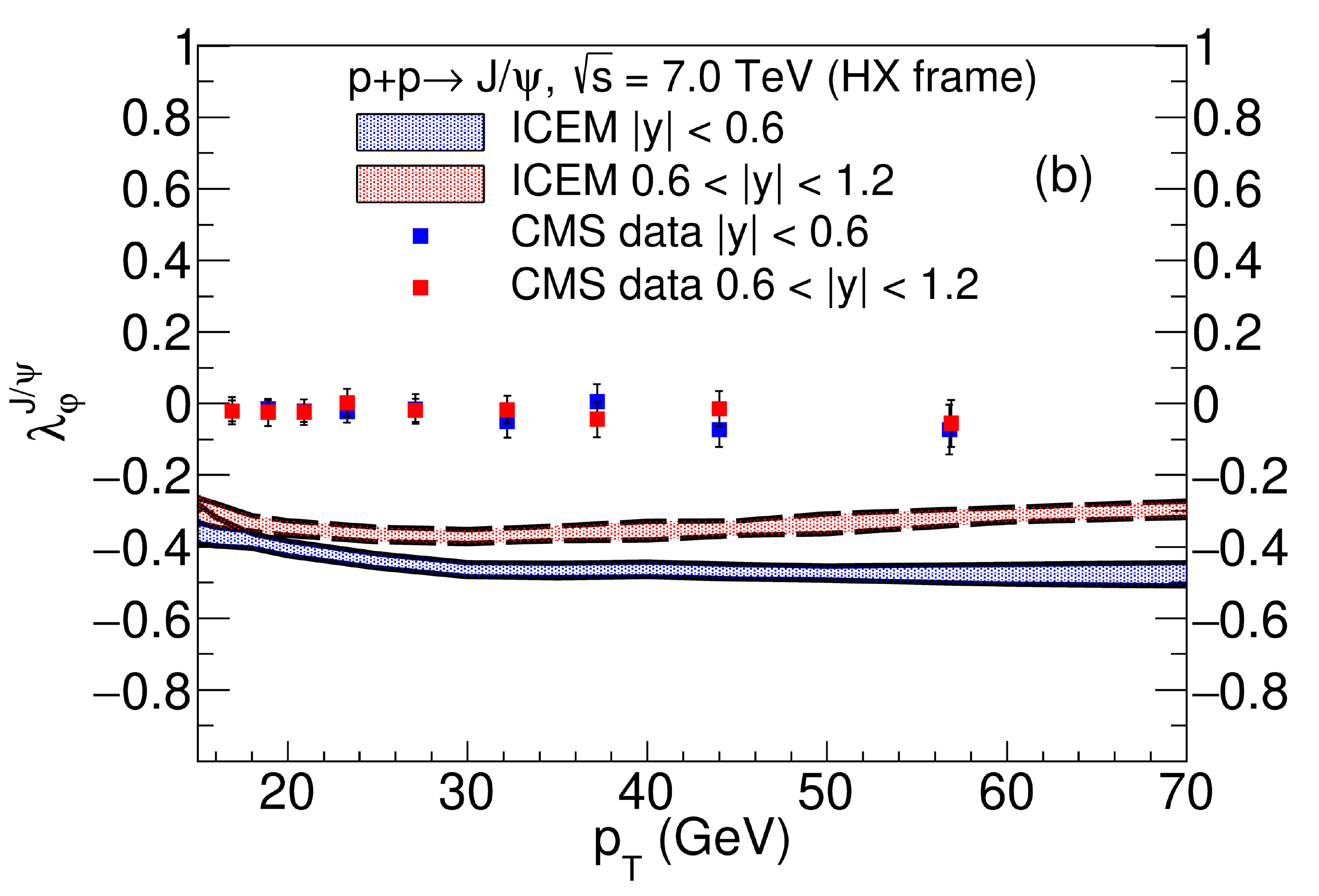}
\end{minipage}
\begin{minipage}[ht]{0.68\columnwidth}
\centering
\includegraphics[width=\columnwidth]{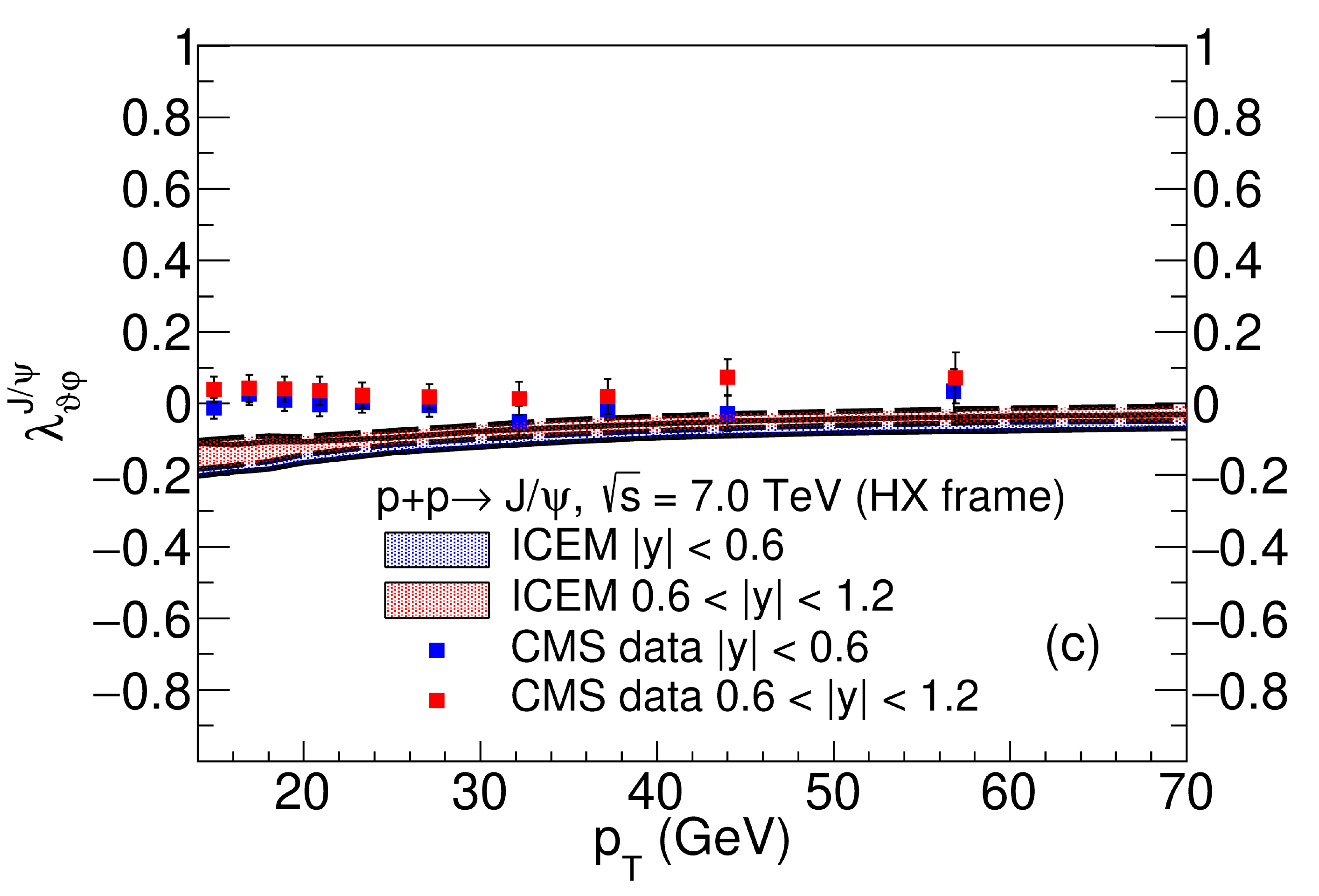}
\end{minipage}
\caption{(a) The polar anisotropy parameter ($\lambda_\vartheta$), (b) the azimuthal anisotropy parameter ($\lambda_\varphi$), and (c) the polar-azimuthal correlation parameter ($\lambda_{\vartheta\varphi}$) in the helicity frame at $\sqrt{s} = 7$~TeV in the ICEM. The combined mass, renormalization scale, and factorization scale uncertainties are shown in the band and compared to the CMS data \cite{Chatrchyan:2013cla} in $|y|<0.6$ (blue) and $0.6<|y|<1.2$ (red).} \label{cms-frame-dependent-lambdas-hx}
\end{figure*}

\begin{figure*}
\centering
\begin{minipage}[ht]{0.68\columnwidth}
\centering
\includegraphics[width=\columnwidth]{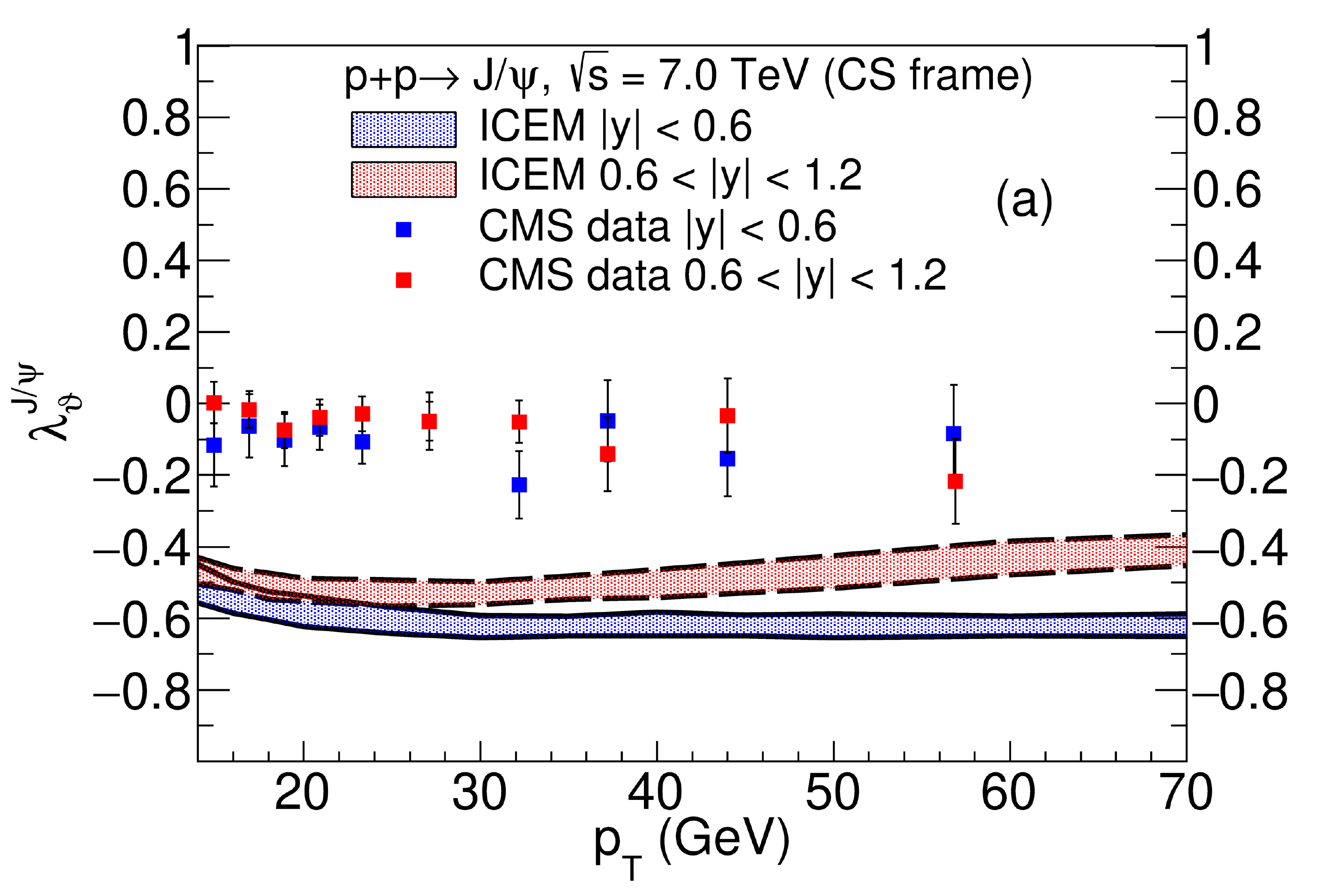}
\end{minipage}%
\begin{minipage}[ht]{0.68\columnwidth}
\centering
\includegraphics[width=\columnwidth]{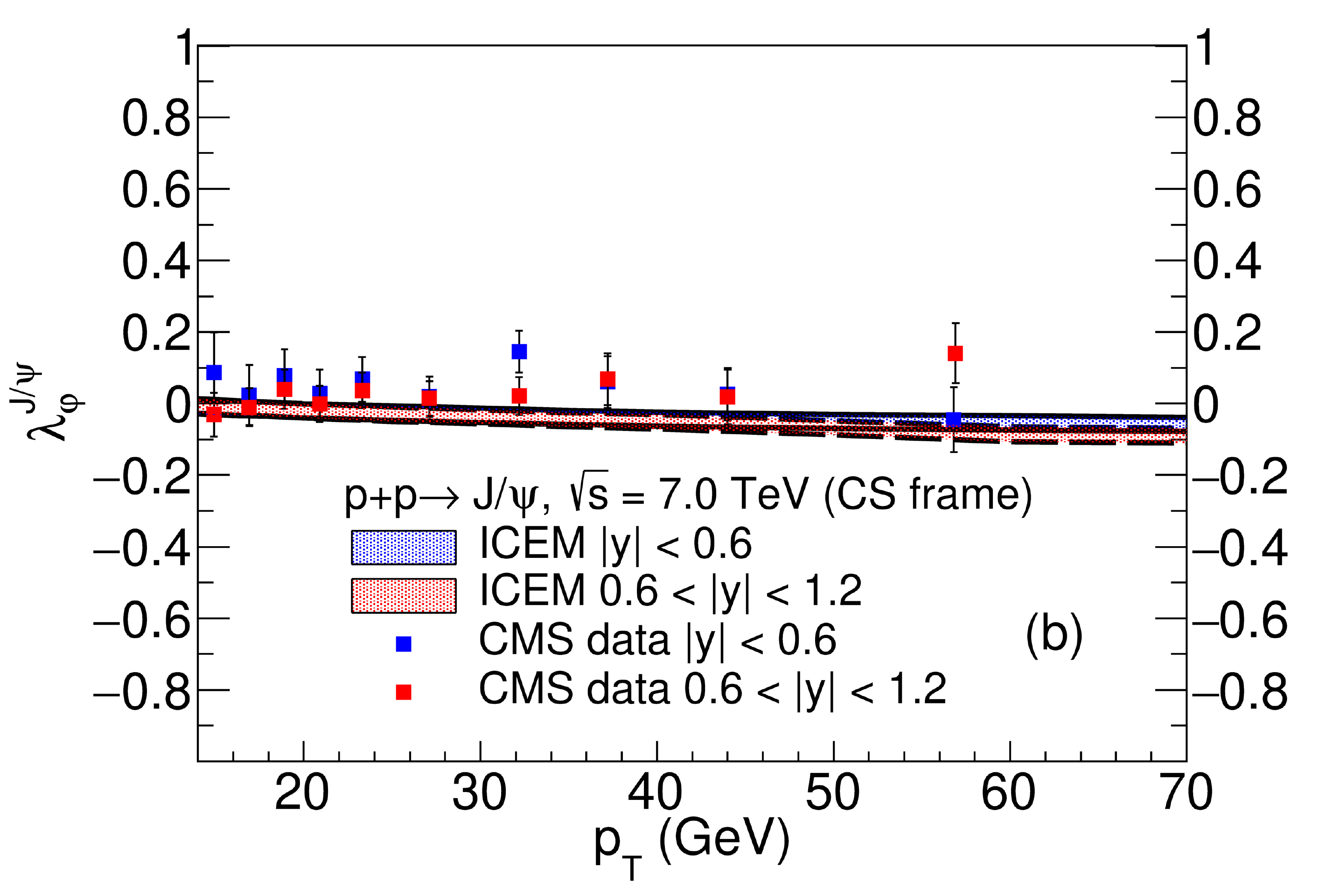}
\end{minipage}
\begin{minipage}[ht]{0.68\columnwidth}
\centering
\includegraphics[width=\columnwidth]{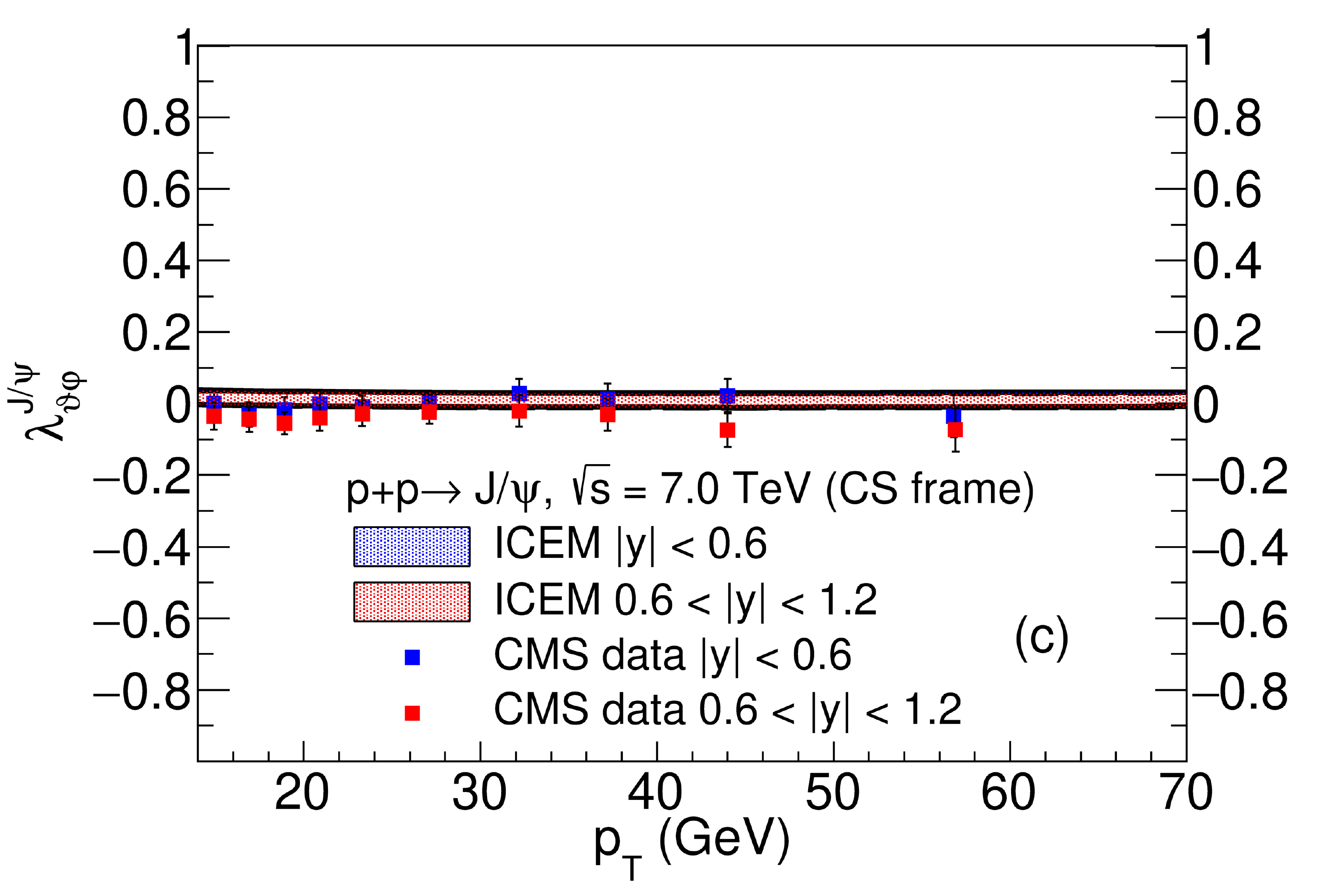}
\end{minipage}
\caption{(a) The polar anisotropy parameter ($\lambda_\vartheta$), (b) the azimuthal anisotropy parameter ($\lambda_\varphi$), and (c) the polar-azimuthal correlation parameter ($\lambda_{\vartheta\varphi}$) in the Collins-Soper frame at $\sqrt{s} = 7$~TeV in the ICEM. The combined mass, renormalization scale, and factorization scale uncertainties are shown in the band and compared to the CMS data \cite{Chatrchyan:2013cla} in $|y|<0.6$ (blue) and $0.6<|y|<1.2$ (red).} \label{cms-frame-dependent-lambdas-cs}
\end{figure*}

\subsection{$p_T$ and $y$ dependences of $\lambda_\vartheta$, $\lambda_{\varphi}$, and $\lambda_{\vartheta\varphi}$}

\begin{figure*}
\centering
\begin{minipage}[ht]{0.68\columnwidth}
\centering
\includegraphics[width=\columnwidth]{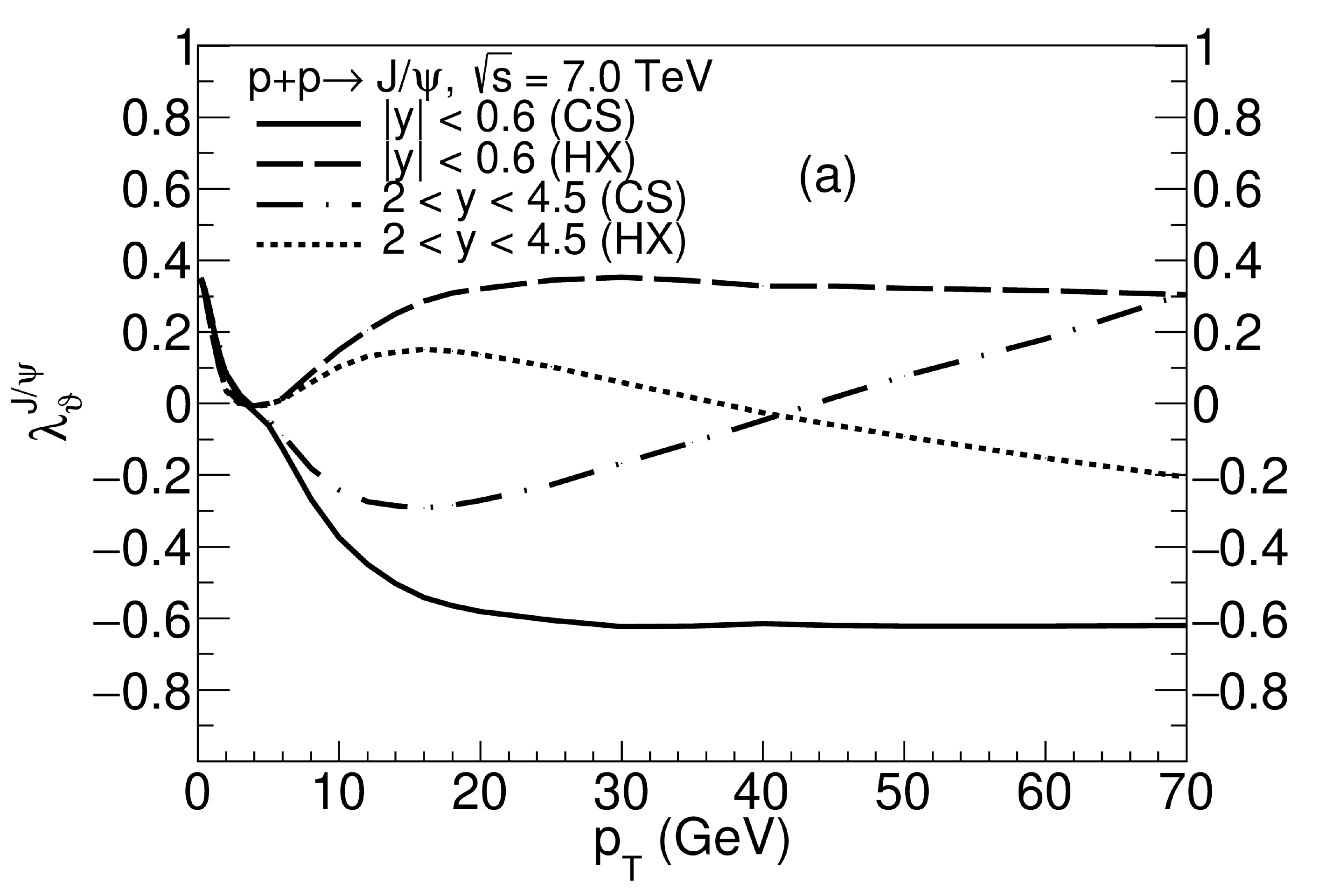}
\end{minipage}%
\begin{minipage}[ht]{0.68\columnwidth}
\centering
\includegraphics[width=\columnwidth]{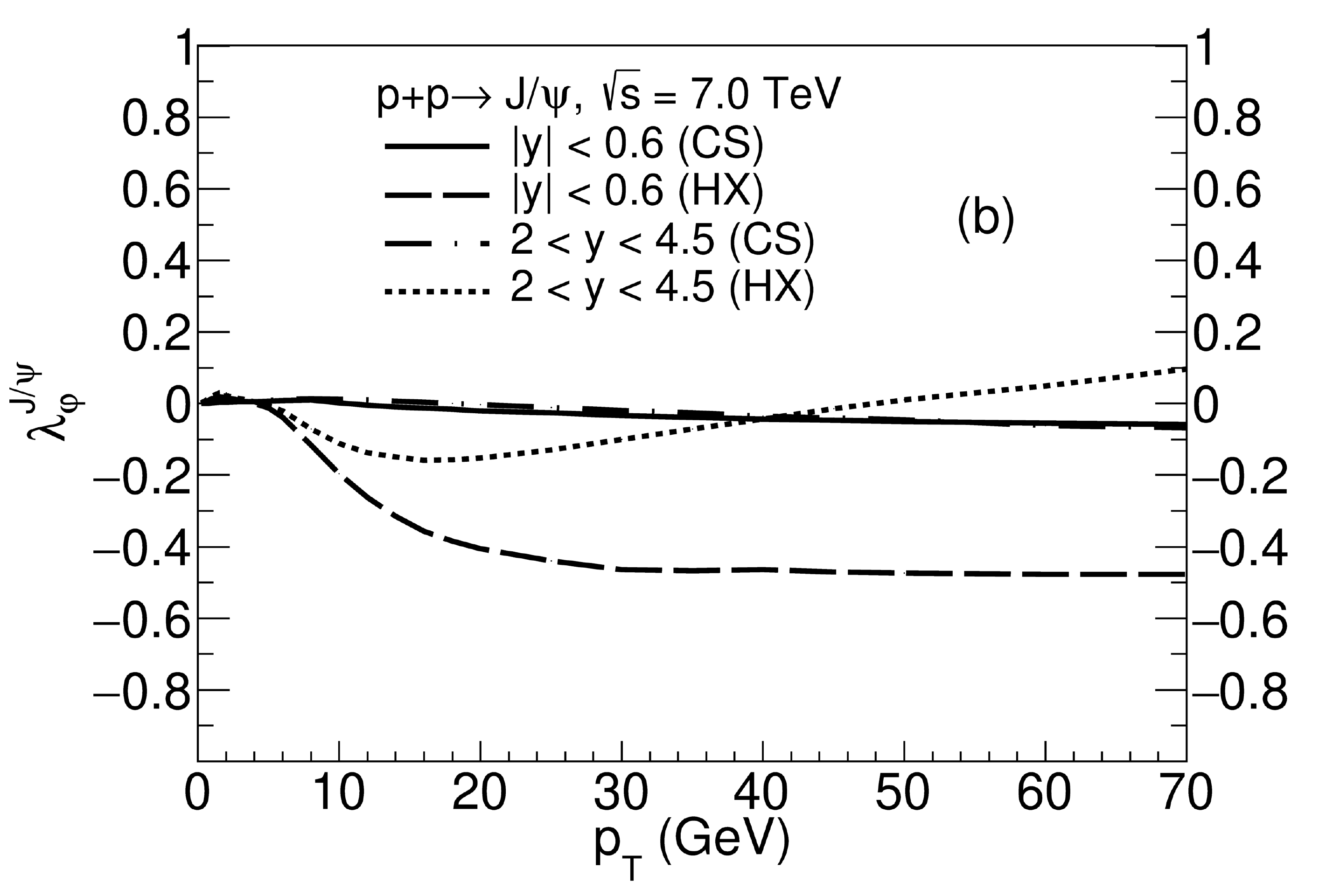}
\end{minipage}
\begin{minipage}[ht]{0.68\columnwidth}
\centering
\includegraphics[width=\columnwidth]{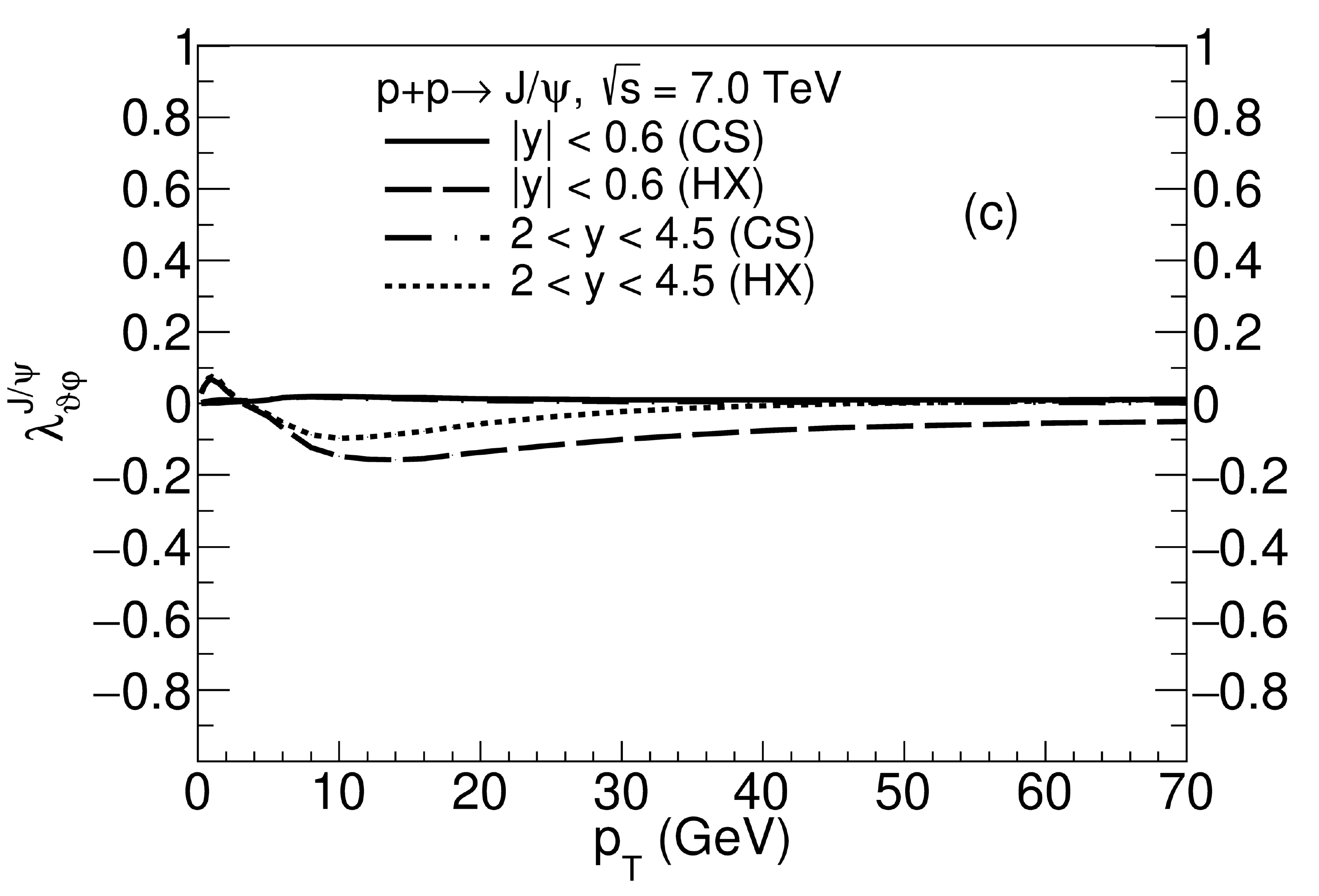}
\end{minipage}
\caption{(a) The polar anisotropy parameter ($\lambda_\vartheta$), (b) the azimuthal anisotropy parameter ($\lambda_\varphi$), and (c) the polar-azimuthal correlation parameter ($\lambda_{\vartheta\varphi}$) at $\sqrt{s} = 7$~TeV in the ICEM. The ICEM results in the Collins-Soper frame in $|y|<0.6$ (solid) and in $2<y<4.5$ (dot-dashed), and the results in the helicity frame in $|y|<0.6$ (dashed) and in $2<y<4.5$ (dotted) are shown. Only the baseline results are shown here.} \label{all-frame-dependent-lambdas}
\end{figure*}

\begin{figure*}
\centering
\begin{minipage}[ht]{0.63\columnwidth}
\centering
\includegraphics[width=\columnwidth]{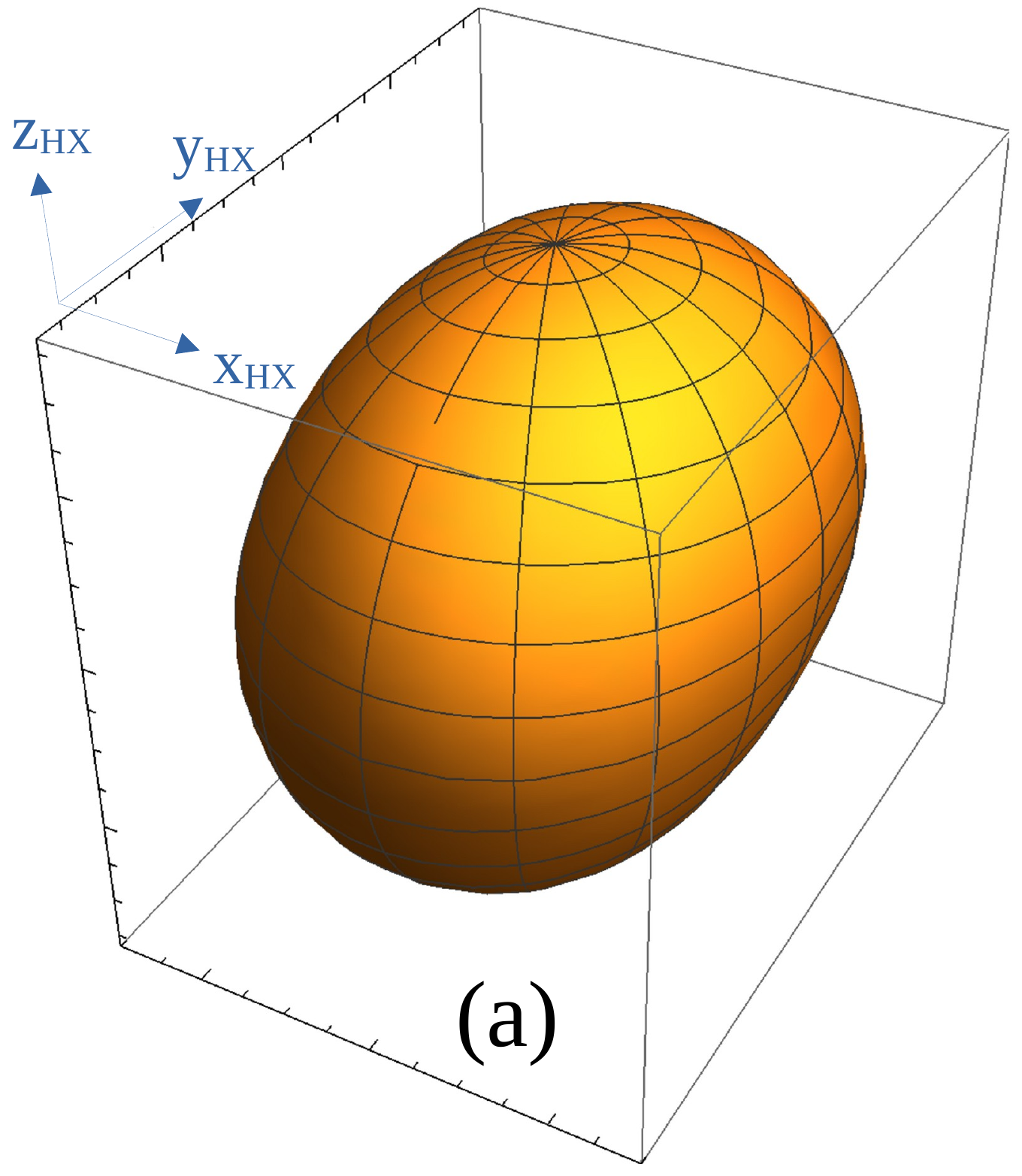}
\end{minipage}%
\begin{minipage}[ht]{0.65\columnwidth}
\centering
\includegraphics[width=\columnwidth]{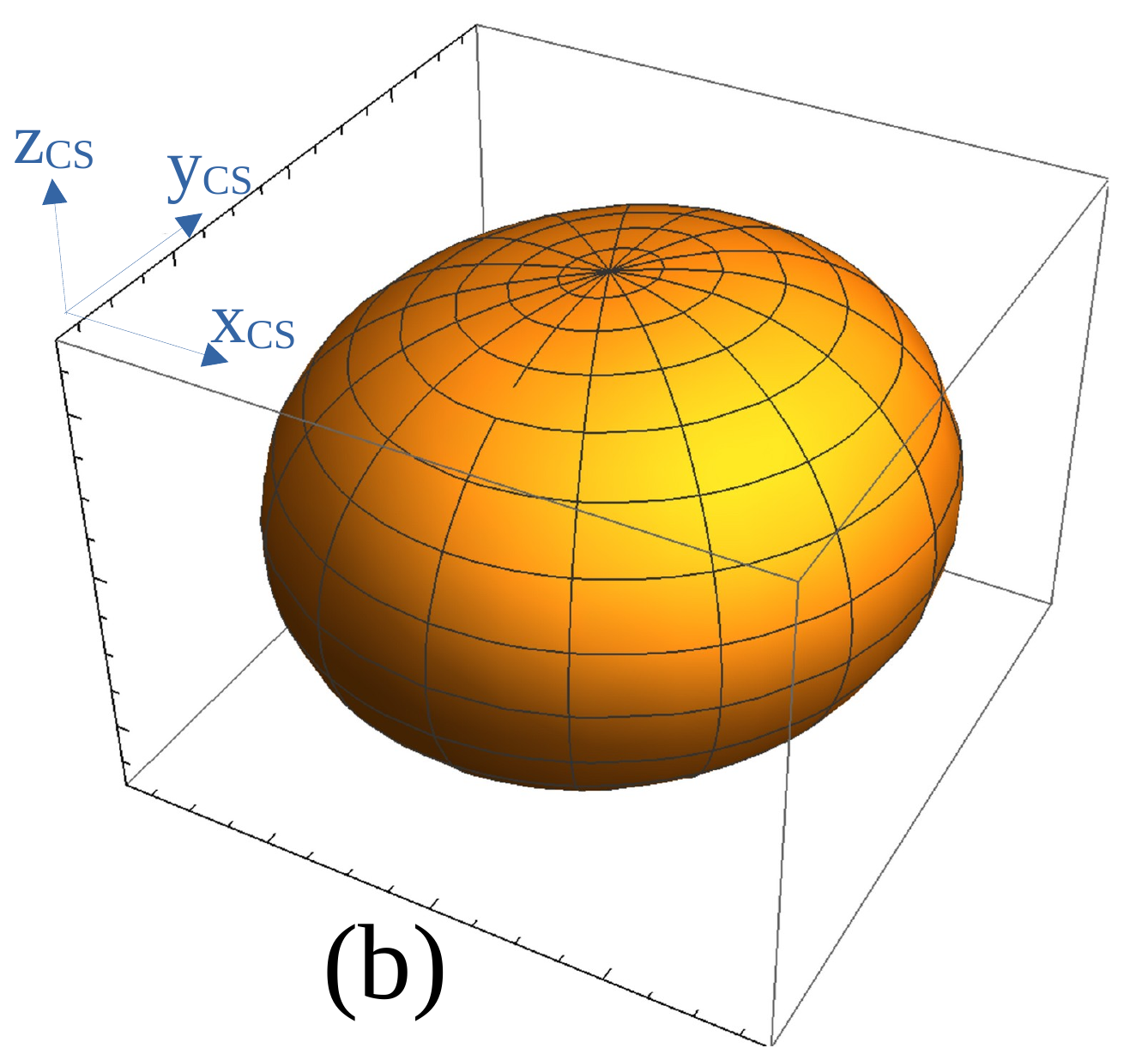}
\end{minipage}
\begin{minipage}[ht]{0.76\columnwidth}
\centering
\includegraphics[width=\columnwidth]{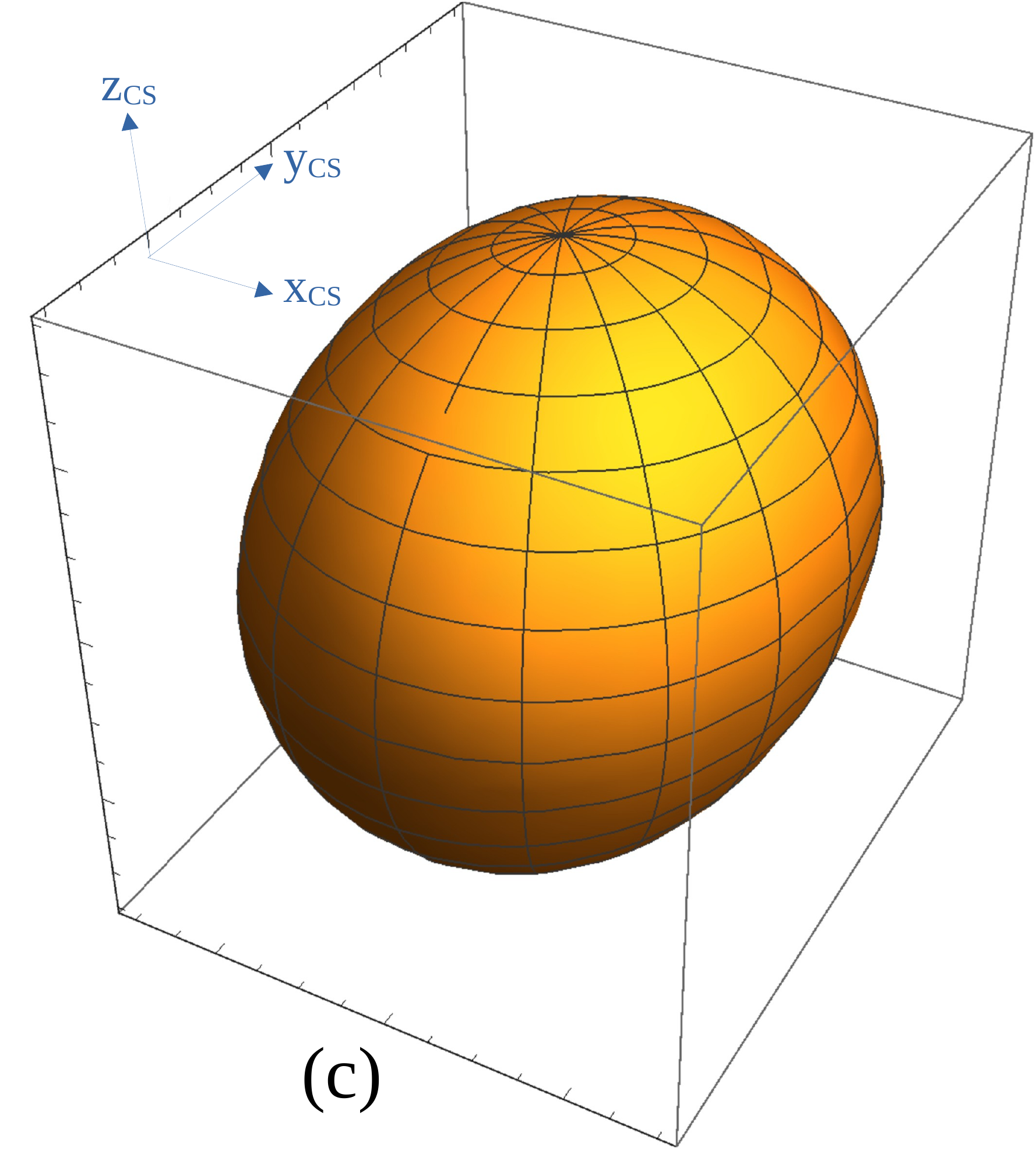}
\end{minipage}
\caption{(a) The angular distribution of the ICEM direct $J/\psi$ production in the Collins-Soper frame and (b) in the helicity frame at $p_T=12$~GeV. They represent the same angular distribution separated by one rotation. We constructed the angular distribution in the Collins-Soper frame based on the data collected in the $10<p_T<15$~GeV bin by the LHCb Collaboration \cite{Aaij:2013nlm}, and it is shown in (c) for comparison.} \label{ch6-angular-distributions}
\end{figure*}

\begin{figure}[h!]
\centering
\includegraphics[width=\columnwidth]{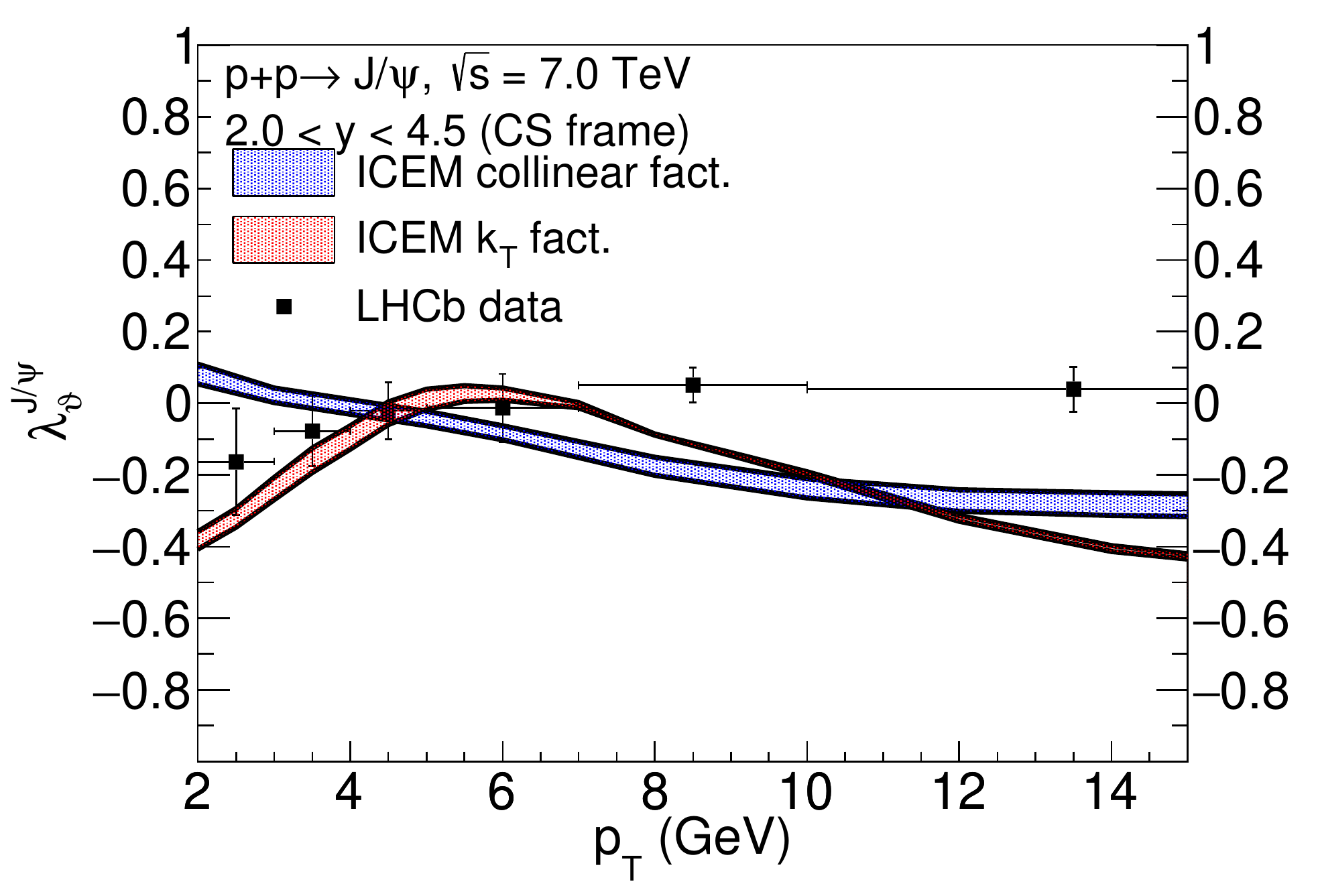}
\caption{The polar anisotropy parameter ($\lambda_\vartheta$) of direct $J/\psi$ production at $\sqrt{s} = 7$~TeV with $2<y<4.5$ in the Collins-Soper frame calculated in the collinear ICEM in this work (blue) and in the $k_T$-factorized ICEM in Ref.~\cite{Cheung:2018tvq}. Both calculations include only the $gg$ contributions and are compared to the LHCb data \cite{Aaij:2013nlm}.} \label{ch6-icem-kt-collinear-compare}
\end{figure}

We calculate the $p_T$ dependence of the frame-dependent polarization parameters $\lambda_\vartheta$, $\lambda_{\varphi}$, and $\lambda_{\vartheta\varphi}$ at $\sqrt{s}=7$~TeV in the helicity frame and in the Collins-Soper frame. We compare the polarization parameters at low and moderate $p_T$ ($2<p_T<15$~GeV) with the data measured by the LHCb Collaboration \cite{Aaij:2013nlm} and the ALICE Collaboration \cite{Abelev:2011md}, where the data are collected in rapidity ranges $2<y<4.5$ and $2.5<y<4$. The comparisons in the helicity frame and in the Collins-Soper frame are presented in Figs.~\ref{frame-dependent-lambdas-hx} and ~\ref{frame-dependent-lambdas-cs}, respectively. We then compare the polarization parameters at high $p_T$ ($14<p_T<70$~GeV) with the data measured by the CMS Collaboration \cite{Abelev:2011md}, where the data are collected in rapidity ranges $|y|<0.6$ and $0.6<|y|<1.2$. These comparisons in the helicity frame and in the Collins-Soper frame are presented in Figs.~\ref{cms-frame-dependent-lambdas-hx} and \ref{cms-frame-dependent-lambdas-cs}, respectively. 

The polar anisotropy parameter ($\lambda_{\vartheta}$) reflects the proportion of the $J/\psi$ in each spin projection state, with $\lambda_{\vartheta} = 1$ referring to completely transverse production of $J_z=\pm1$ and $\lambda_{\vartheta} = -1$ referring to completely longitudinal production of $J_z=0$. At low $p_T$, $\lambda_{\vartheta}$ is close to zero in both the helicity frame and the Collins-Soper frame, indicating equal $J/\psi$ yields in each spin projection state ($J_z = 0$, $\pm1$). However, as $p_T$ grows larger, the difference between the $\lambda_{\vartheta}$ calculated in the two frames increases with $p_T$. In the helicity frame, the transverse component ($J_z = \pm1$) falls off more slowly than the longitudinal component ($J_z = 0$). As a result, $\lambda_{\vartheta}$ becomes positive as $p_T$ increases. This is consistent with the (color glass condensate) CGC$+$NRQCD approach at low and moderate $p_T$ \cite{Ma:2018qvc} and a NRQCD calculation at high $p_T$ \cite{Chao:2012iv}. On the other hand, in the Collins-Soper frame, the longitudinal component dominates with increasing $p_T$. Thus, $\lambda_{\vartheta}$ becomes negative as $p_T$ increases. This relative behavior of $\lambda_{\vartheta}$ in the two frames is expected because the polarization $z$ axes are parallel at $p_T = 0$ and become orthogonal in the limit $p_T\rightarrow \infty$.

\begin{figure*}
\centering
\begin{minipage}[ht]{0.68\columnwidth}
\centering
\includegraphics[width=\columnwidth]{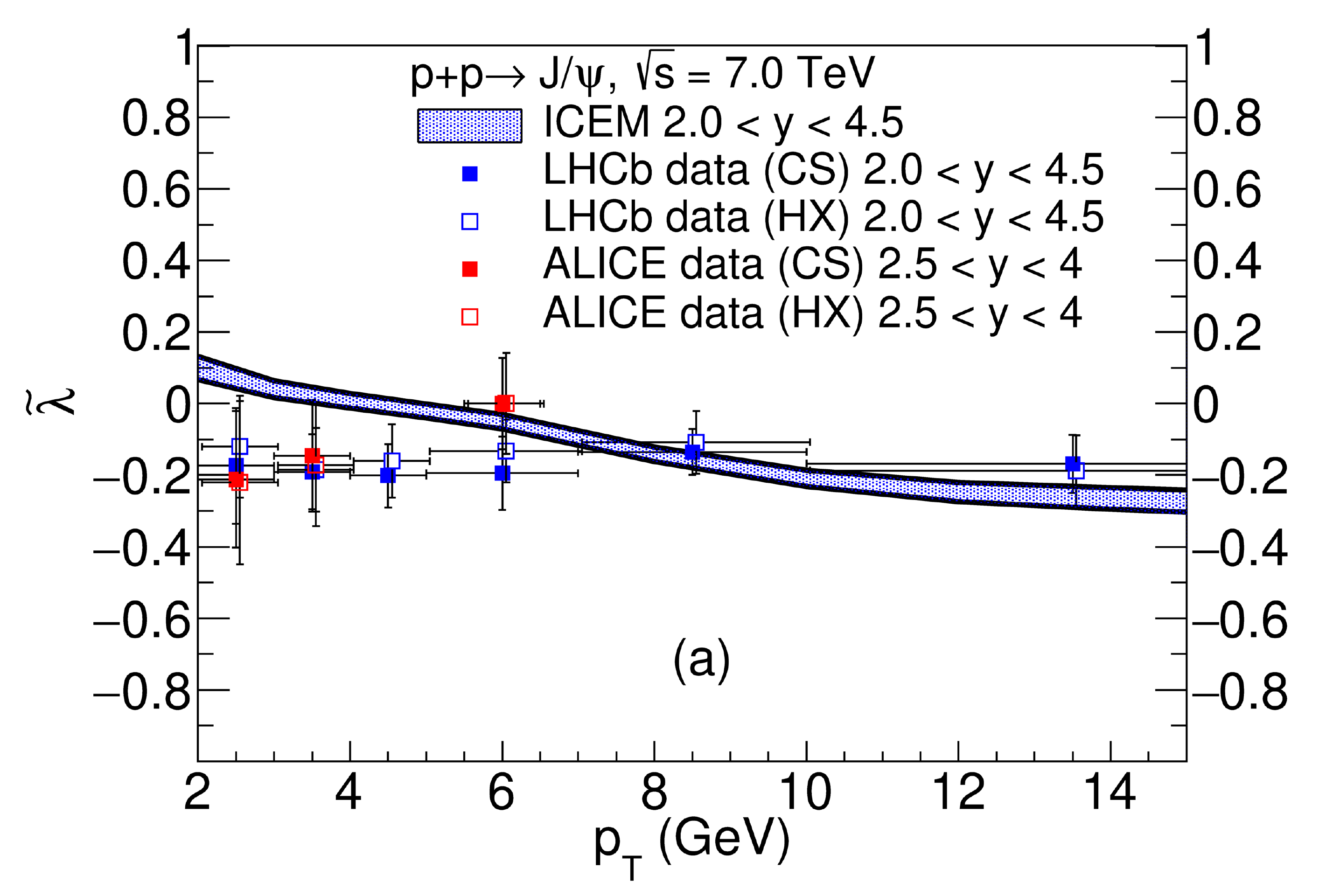}
\end{minipage}%
\begin{minipage}[ht]{0.68\columnwidth}
\centering
\includegraphics[width=\columnwidth]{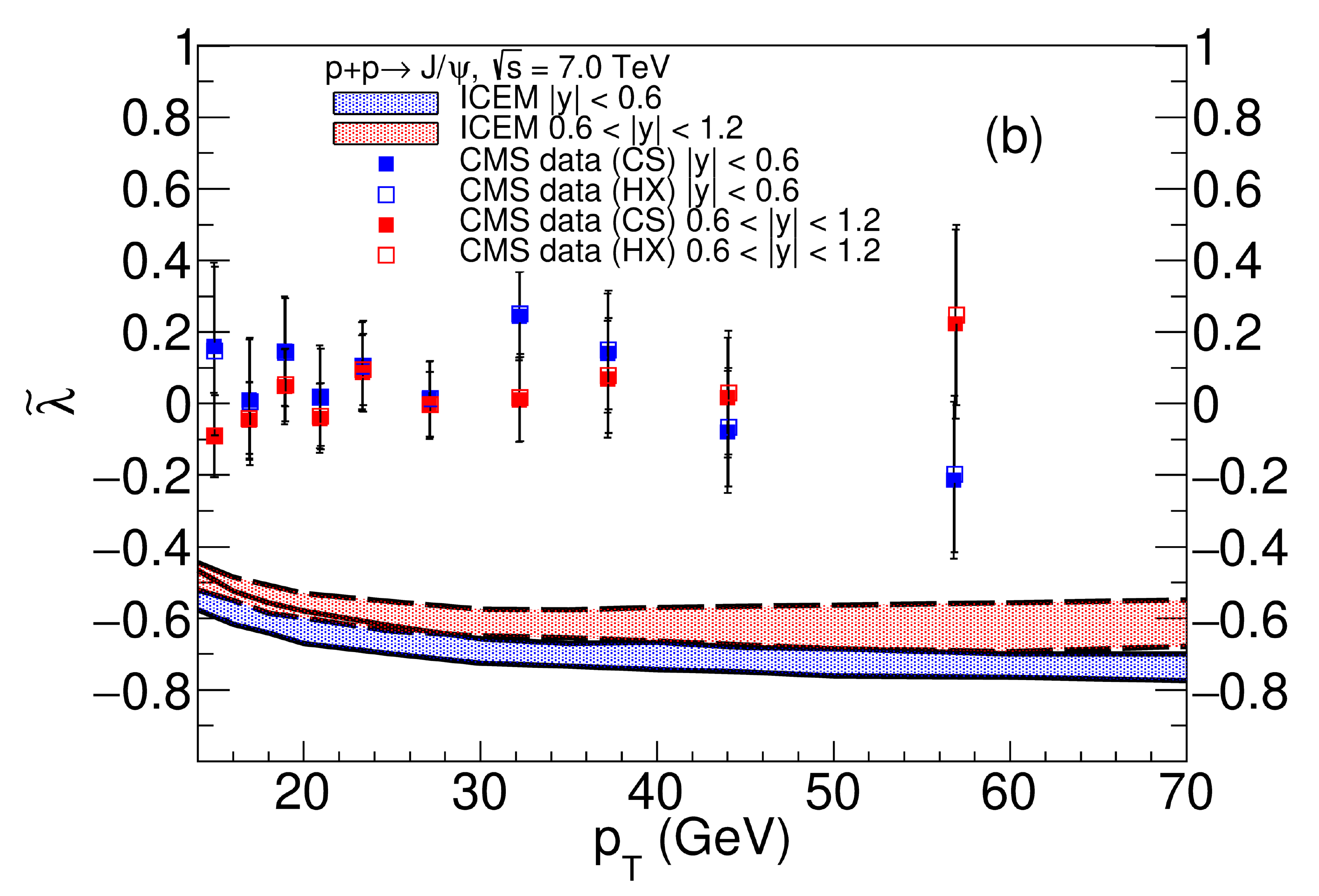}
\end{minipage}
\begin{minipage}[ht]{0.68\columnwidth}
\centering
\includegraphics[width=\columnwidth]{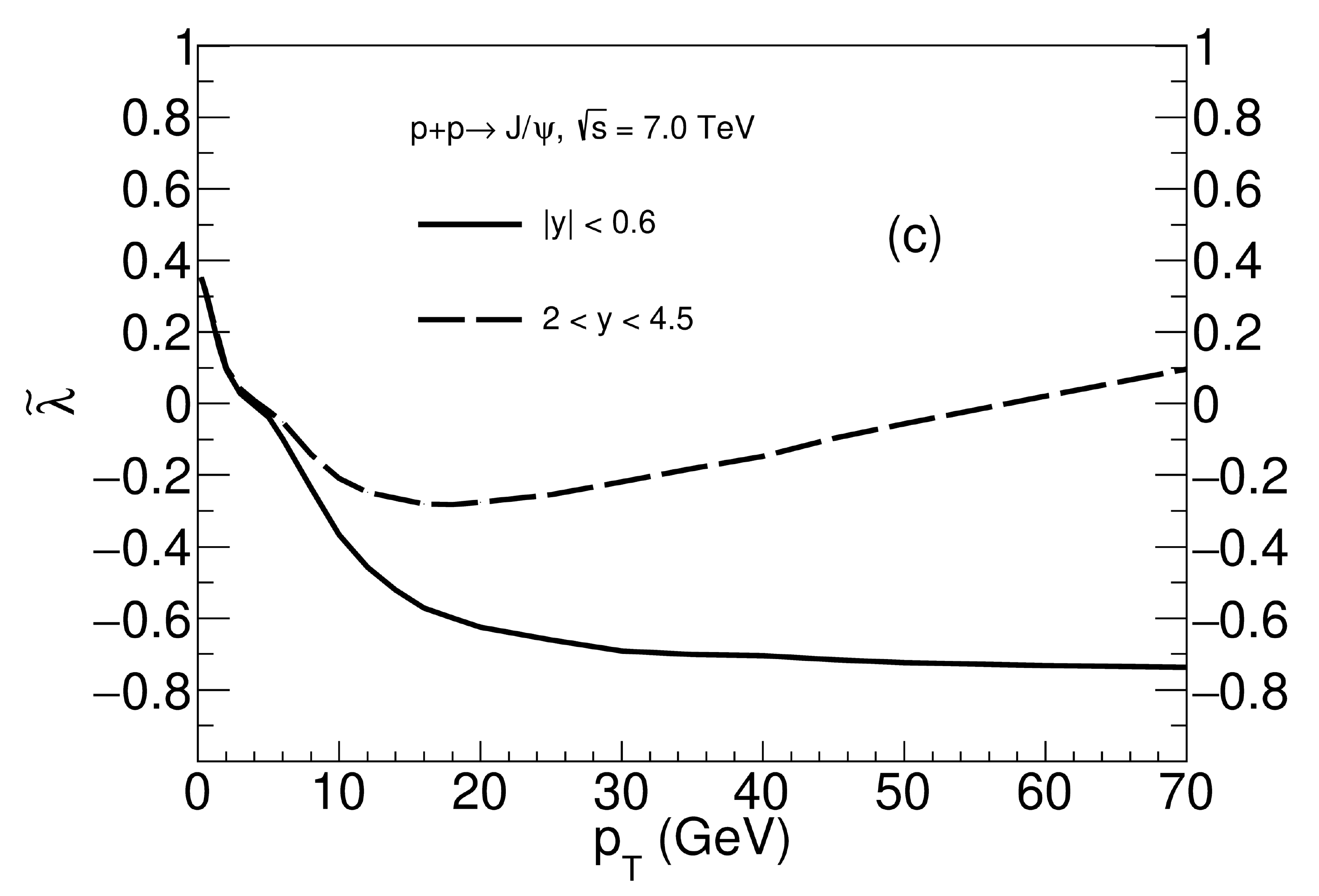}
\end{minipage}%
\caption{The $p_T$ dependence of the frame-invariant polarization parameter, $\tilde{\lambda}$, in the ICEM (a) compared to LHCb data \cite{Aaij:2013nlm} (blue) and the ALICE data \cite{Abelev:2011md} (red) at low $p_T$; (b) compared to CMS data \cite{Chatrchyan:2013cla} in the most central rapidity bin (blue) and in the next rapidity bin (red) at high $p_T$; and (c) the baseline ICEM results in $|y|<0.6$ (solid) and $2<y<4.5$ (dashed). Data measured in the Collins-Soper frame are presented as solid points, and those in the helicity frame are presented as open points. The data in the helicity frame are displaced by 0.05~GeV for visualization purposes. } \label{ch6-lambda-invariant}
\end{figure*}

The azimuthal anisotropy parameter ($\lambda_\varphi$) reflects the azimuthal symmetry of $J/\psi$ production. When $\lambda_\varphi=0$, the production is azimuthally symmetric. When $\lambda_\varphi=\pm1$, the azimuthal distribution is maximally asymmetric. We note that this parameter strongly depends on the production mechanism as well as the frame the distribution is measured in. In the Collins-Soper frame, this parameter is close to zero over all $p_T$ as the matrix element $\sigma_{+1,-1}$ is small relative to $\sigma_{+1,+1}$ and $\sigma_{0,0}$. This means that the $z_{CS}$ axis is approximately the azimuthal symmetry axis. In the helicity frame, the matrix element $\sigma_{+1,-1}$ is negative and become more negative. As a result, $\lambda_\varphi$ becomes negative as $p_T$ grows larger, showing that the $z_{HX}$ axis is not the symmetry axis of the distribution. However, the distribution itself is rotationally invariant. The discrepancy between $\lambda_\varphi$ in these two frames is a combination of two factors: $z_{CS}$ and $z_{HX}$ becomes approximately orthogonal as $p_T$ increases and production is not spherically symmetric.

The polar-azimuthal correlation parameter ($\lambda_{\vartheta\varphi}$) describes the angular correlation between $2\vartheta$ and $\varphi$. When $\lambda_{\vartheta\varphi}=0$, the two angles are uncorrelated and as $\lambda_{\vartheta\varphi}$ departs from 0, the behavior of the distribution becomes similar at locations where $2\vartheta=\varphi$. In both the helicity frame and the Collins-Soper frame, $\lambda_{\vartheta\varphi}$ is consistent with~0 at forward rapidity, which agrees with the data. At central rapidity, $\lambda_{\vartheta\varphi}$ pulls away from 0, but in general within the uncertainties of the data.

We observe a substantial rapidity dependence in the polarization parameters. To illustrate the rapidity dependence, we present the polarization parameters as a function of $p_T$ in Fig.~\ref{all-frame-dependent-lambdas} in the most central rapidity and the forward rapidity regions covered by the CMS detector and the LHCb detector respectively. In both the Collins-Soper frame and in the helicity frame, the polar anisotropy parameter, $\lambda_{\vartheta}$ switches sign in the high $p_T$ limit as rapidity increases. We observe a difference in the curvature of the polarized cross sections as a function of $p_T$ in the forward rapidity region compared to the central rapidity region. At central rapidity, $\lambda_{\vartheta}$ becomes constant for $p_T\gtrsim20$~GeV as the longitudinal and transverse cross sections fall off at similar rates while keeping an approximately fixed ratio. At forward rapidity, in the Collins-Soper frame, the longitudinal cross section first falls off faster than the transverse cross section at moderate $p_T$, but at high $p_T$, the decrease of the longitudinal cross section slows down until it eventually dominates production. We observe the opposite behavior in the helicity frame, where the transverse cross section falls off more slowly than the longitudinal cross section and ultimately dominates production. 

We also observe a rapidity dependence in $\lambda_{\varphi}$. At central rapidity, the polarized cross section matrix element, $\sigma_{+1,-1}$, from the $gg$ contribution in the helicity frame is negative and relatively constant for $p_T\gtrsim20$~GeV. As rapidity increases, the matrix element becomes smaller at moderate $p_T$ and finally becomes small and positive at high $p_T$. Thus, $\lambda_\varphi$ follows the $gg$ contribution, and the production becomes more azimuthally symmetric with increasing rapidity. In the Collins-Soper frame, the azimuthal dependence of the distribution shows the opposite behavior as rapidity increases. Compared to at central rapidity, the magnitude of $\lambda_{\varphi}$ at forward rapidity is larger at both moderate $p_T$ and high $p_T$. As a result, the production becomes less azimuthally symmetric as rapidity increases. $\lambda_{\vartheta\varphi}$ is also rapidity dependent because the magnitude of the matrix element, $\sigma_{0,+1}$ in both frames is smaller at forward rapidity than at central rapidity. Note that when the polarization parameters are close to zero, such as $\lambda_{\varphi}$ and $\lambda_{\vartheta\varphi}$, the variation with rapidity is harder to visualize because a given fractional change in the size of a matrix element corresponds to a smaller change in the polarization parameter.

The resulting angular distributions in the helicity frame and in the Collins-Soper frame at $p_T = 12$~GeV are shown in Figs.~\ref{ch6-angular-distributions}(a) and \ref{ch6-angular-distributions}(b), respectively. Note that two distributions are almost identical except that one appears to be rotated 90$^{\circ}$ with respect to the other. Thus, the two distributions can be interpreted as a top view and a side view of the production distributions. The angular distribution in the Collins-Soper frame, based on the data collected in the $10<p_T<15$~GeV bin by the LHCb Collaboration \cite{Aaij:2013nlm}, is presented in Fig.~\ref{ch6-angular-distributions}(c).

We compare the polar anisotropy parameter, $\lambda_\vartheta$, in this calculation with the one calculated in the polarized ICEM using the $k_T$-factorization approach \cite{Cheung:2018tvq} in Fig.~\ref{ch6-icem-kt-collinear-compare}. Since the $k_T$-factorized ICEM considers only the contribution from off-shell Reggeized gluons at $\mathcal{O}(\alpha_s^2)$, we compare the $p_T$ dependence of $\lambda_\vartheta$ in the $k_T$-factorized ICEM with that in this calculation using the contribution from $gg\rightarrow c\bar{c}g$ at $\mathcal{O}(\alpha_s^3)$ only. The polarization calculated in the $gg$ channel alone is approximately equivalent to that summed over all channels in the collinear factorization approach because the $gg$ component dominates both the polarized and unpolarized cross sections. Although the kinematics in the two calculations are different, they both predict longitudinal polarization at high $p_T$, where the experimental results are nearly unpolarized. At small and moderate $p_T$, the difference in $\lambda_\vartheta$ is within the uncertainty of the data.

\subsection{$p_T$ and $y$ dependence of $\tilde{\lambda}$}

In Figs.~\ref{cms-frame-dependent-lambdas-hx} and \ref{cms-frame-dependent-lambdas-cs}, we observe that, at high $p_T$, the ICEM is in better agreement with the measured data in the helicity frame than in the Collins-Soper frame. However, even though we are switching from one frame to another, we are still comparing the same angular distributions. The difference between the ICEM polarization results and the data is then best quantified by a frame-independent polarization parameter. Thus, we compute the frame-invariant polarization parameter $\tilde{\lambda}$ as a function of $p_T$ using $\lambda_\vartheta$ and $\lambda_\varphi$. We compare $\tilde{\lambda}$ as a function of $p_T$ with the data from LHCb and CMS in the helicity and the Collins-Soper frames in Fig.~\ref{ch6-lambda-invariant}.

Since the azimuthal anisotropy parameter $\lambda_\varphi$ in the Collins-Soper frame is close to zero in all $p_T$ and $y$ ranges considered, the $p_T$ and $y$ dependences of the invariant polarization parameter $\tilde{\lambda}$ are very similar to those in $\lambda_\vartheta$ of the Collins-Soper frame. We find the curves are generally within $1\sigma$ of the low and moderate $p_T$ data, as shown in Fig.~\ref{ch6-lambda-invariant}~(a). We also find the invariant polarization of direct $J/\psi$ is relatively constant at high $p_T$. We observe a rapidity dependence similar to that of the $\lambda_{\vartheta}$ in the high $p_T$ limit, where the invariant polarization is less longitudinal in the forward rapidity region than in the central rapidity region. We note that the invariant polarization parameter was devised as a way to treat each frame equally rather than to be the final arbiter between theoretical model calculations. Our direct $J/\psi$-invariant polarization results are in reasonable agreement with the measured data over all $p_T$ even though our calculations show a decrease in $\tilde{\lambda}$ with $p_T$. We note that including feed down from higher mass states could change the slope in this $p_T$ range. Our results are in agreement with the $J/\psi$-invariant polarization found in the CGC$+$NRQCD approach at low and moderate $p_T$ \cite{Ma:2018qvc}.

\section{Conclusions}

We have presented the transverse-momentum dependence of the direct $J/\psi$ cross section as well as the polarization in $p+p$ collisions in the improved color evaporation model in the collinear factorization approach. We compare the $p_T$ and $y$ dependences to data for inclusive $J/\psi$ production measured by the LHCb, ALICE, and CMS Collaboration. We also present the frame-invariant parameter, $\tilde{\lambda}$, as a function of $p_T$ and compare our results with the data. We find direct $J/\psi$ production is consistent with the unpolarized data at small and moderate $p_T$ and becomes longitudinal in the high $p_T$ limit. In the near future we will see if the feed-down contribution will have a positive influence on the parameter to match the data in that region. We will study the effects of feed-down production in this approach in a future publication.


\section{Acknowledgments}
This work was performed under the auspices of the U.S. Department of Energy by Lawrence Livermore National Laboratory under Contract No. DE-AC52-07NA27344 and supported by the U.S. Department of Energy, Office of Science, Office of Nuclear Physics (Nuclear Theory) under Contract No. DE-SC-0004014, and the LLNL-LDRD Program under Project No. 21-LW-034.


\end{document}